\documentclass{pasa}%
\jid{PASA}
\doi{10.1017/pas.\the\year.xxx}
\jyear{\the\year}

\usepackage[authoryear]{natbib}
\usepackage{url}
\usepackage{caption}
\usepackage{subcaption}
\usepackage{threeparttable}
\bibpunct{(}{)}{;}{a}{}{,}
\title[The GLEAM survey]{GLEAM: The GaLactic and Extragalactic All-sky MWA survey.}
\author[Wayth et al.]{R.~B.~Wayth$^{1,2}$,
E.~Lenc$^{3,2}$,
M.~E.~Bell$^{4,2}$,
J.~R.~Callingham$^{3,2,4}$,
K.~S.~Dwarakanath$^{5}$,
T.~M.~O.~Franzen$^{1}$,
B.-Q.~For$^{6}$,
B.~Gaensler$^{3,7,2}$,
P.~Hancock$^{1,2}$,
L.~Hindson$^{8}$,
N.~Hurley-Walker$^{1}$,
C.~A.~Jackson$^{1,2}$,
M.~Johnston-Hollitt$^{8}$,
A.~D.~Kapi\'{n}ska$^{6,2}$,
B.~McKinley$^{9,2}$,
J.~Morgan$^{1}$,
A.~R.~Offringa$^{10}$,
P.~Procopio$^{9,2}$,
L.~Staveley-Smith$^{6,2}$,
C.~Wu$^{6}$,
Q.~Zheng$^{8}$,
C.~M.~Trott,$^{1,2}$
G.~Bernardi$^{11,12,13}$,
J.~D.~Bowman$^{14}$, 
F.~Briggs$^{15}$,
R.~J.~Cappallo$^{16}$,
B.~E.~Corey$^{16}$,
A.~A.~Deshpande$^{5}$,
D.~Emrich$^{1}$,
R.~Goeke$^{17}$,
L.~J.~Greenhill$^{13}$,
B.~J.~Hazelton$^{18}$,
D.~L.~Kaplan$^{19}$,
J.~C.~Kasper$^{13,20}$,
E.~Kratzenberg$^{16}$,
C.~J.~Lonsdale$^{16}$,
M.~J.~Lynch$^{1}$,
S.~R.~McWhirter$^{16}$,
D.~A.~Mitchell$^{4,2}$,
M.~F.~Morales$^{18}$,
E.~Morgan$^{17}$,
D.~Oberoi$^{21}$,
S.~M.~Ord$^{1,2}$,
T.~Prabu$^{5}$,
A.~E.~E.~Rogers$^{16}$, 
A.~Roshi$^{22}$,
N.~Udaya~Shankar$^{5}$, 
K.~S.~Srivani$^{5}$, 
R.~Subrahmanyan$^{5,2}$, 
S.~J.~Tingay$^{1,2}$,
M.~Waterson$^{1}$,
R.~L.~Webster$^{9,2}$,
A.~R.~Whitney$^{16}$,
A.~Williams$^{1}$,
C.~L.~Williams$^{17}$\\
\\
\affil{$^1$International Centre for Radio Astronomy Research (ICRAR), Curtin University, Bentley, WA 6102, Australia}
\affil{$^2$ARC Centre of Excellence for All-Sky Astrophysics (CAASTRO)}
\affil{$^3$Sydney Institute for Astronomy (SIfA), School of Physics, The University of Sydney, NSW 2006, Australia}
\affil{$^4$CSIRO Astronomy and Space Science (CASS), Marsfield, NSW 2122, Australia}
\affil{$^{5}$Raman Research Institute, Bangalore 560080, India}
\affil{$^{6}$International Centre for Radio Astronomy Research (ICRAR), University of Western Australia, Crawley, WA 6009, Australia}
\affil{$^{7}$Dunlap Institute for Astronomy \& Astrophysics, University of Toronto, 50 St George St, Toronto, ON, M5S 3H4, Canada}
\affil{$^{8}$School of Chemical \& Physical Sciences, Victoria University of Wellington, Wellington 6140, New Zealand}
\affil{$^{9}$School of Physics, The University of Melbourne, Parkville, VIC 3010, Australia}
\affil{$^{10}$Netherlands Institute for Radio Astronomy (ASTRON), PO Box 2, 7990 AA Dwingeloo, The Netherlands}
\affil{$^{11}$Square Kilometre Array South Africa (SKA SA), 3rd Floor, The Park, Park Road, Pinelands, 7405, South Africa}
\affil{$^{12}$Department of Physics and Electronics, Rhodes University, PO Box 94, Grahamstown, 6140, South Africa}
\affil{$^{13}$Harvard-Smithsonian Center for Astrophysics, Cambridge, MA 02138, USA}
\affil{$^{14}$School of Earth and Space Exploration, Arizona State University, Tempe, AZ 85287, USA}
\affil{$^{15}$Research School of Astronomy and Astrophysics, Australian National University, Canberra, ACT 2611, Australia}
\affil{$^{16}$MIT Haystack Observatory, Westford, MA 01886, USA}
\affil{$^{17}$Kavli Institute for Astrophysics and Space Research, Massachusetts Institute of Technology, Cambridge, MA 02139, USA}
\affil{$^{18}$Department of Physics, University of Washington, Seattle, WA 98195, USA}
\affil{$^{19}$Department of Physics, University of Wisconsin--Milwaukee, Milwaukee, WI 53201, USA}
\affil{$^{20}$Department of Atmospheric, Oceanic and Space Sciences, University of Michigan, Ann Arbor, MI, USA}
\affil{$^{21}$National Centre for Radio Astrophysics, Tata Institute for Fundamental Research, Pune 411007, India}
\affil{$^{22}$National Radio Astronomy Observatory, Green Bank, WV, USA}
}

\begin{document}%
\begin{abstract}
GLEAM, the GaLactic and Extragalactic All-sky MWA survey, is a survey of the entire radio sky south of declination $+25^{\circ}$ at frequencies between 72 and 231\,MHz, made with the Murchison Widefield Array (MWA) using a drift scan method that makes efficient use of the MWA's very large field-of-view.
We present the observation details, imaging strategies and theoretical sensitivity for GLEAM.
The survey ran for two years, the first year using 40\,kHz frequency resolution and 0.5\,s time resolution;
the second year using 10\,kHz frequency resolution and 2\,s time resolution.
The resulting image resolution and sensitivity depends on observing frequency, sky pointing and image weighting scheme.
At 154\,MHz the image resolution is approximately $2.5 \times 2.2/\cos(\delta+26.7^{\circ})$~arcmin with sensitivity to structures up to $\sim10^{\circ}$ in angular size.
We provide tables to calculate the expected thermal noise for GLEAM mosaics depending on pointing and frequency and discuss limitations to achieving theoretical noise in Stokes I images.
We discuss challenges, and their solutions, that arise for GLEAM including ionospheric effects on source positions and linearly polarised emission, and the instrumental polarisation effects inherent to the MWA's primary beam.

\end{abstract}
\maketitle%
\section{Introduction}
\label{sec:intro}

Low-frequency radio astronomy is once again on the scientific frontier, driven in large part by the goal of measuring the radio emission from high-redshift neutral hydrogen during the Epoch of Reionisation (EoR), predicted to lie in the $50-200$\,MHz part of the radio spectrum \citep[e.g.][]{2006PhR...433..181F,2010ARA&A..48..127M}.
In addition to established telescopes working at these frequencies, such as the Giant Metrewave Radio Telescope \citep[`GMRT',][]{1991CuSc...60...95S}, the Mauritius Radio Telescope \citep{1998JApA...19...35G} and the Jansky Very Large Array \citep[`JVLA',][]{2015AAS...22531104C}, several new telescopes are now operating in this frequency range including the Low Frequency Array \citep[`LOFAR', ][]{2013A&A...556A...2V},
the Precision Array for Probing the Epoch of Reionization \citep[`PAPER', ][]{2010AJ....139.1468P},
the Long Wavelength Array \citep[`LWA', ][]{2013ITAP...61.2540E}, and the Murchison Widefield Array \citep[`MWA',][]{2009IEEEP..97.1497L,2013PASA...30....7T}.
All of these instruments have performed, or are undertaking, large-area sky surveys \citep{2007AJ....134.1245C,Heald15,2014MNRAS.440..327L}\footnote{Also: \url{http://tgss.ncra.tifr.res.in } }. 
The survey properties, including limiting flux density, resolution, surface brightness sensitivity and sky coverage vary considerably between instruments and survey programs.

We present the details of the GaLactic Extragalactic All-sky MWA (GLEAM) survey, which is surveying the sky south of declination $+25^{\circ}$ with the MWA.
GLEAM was conceived to provide data for many science goals including studies of: radio galaxies and active galactic nucleii; galaxy clusters; the Magellanic clouds; diffuse galactic emission and the Galactic magnetic field; galactic and extragalactic spectral lines; supernova remnants; Galactic \textsc{Hii} regions; pulsars and pulsar wind nebulae; and cosmic rays.
GLEAM is intended to leave a significant legacy dataset from the MWA that can be utilised for its capabilities in the time, frequency, polarisation and field-of-view domains.
GLEAM's two year observing program began in August 2013.

The myriad MWA science opportunities enabled by large sky area surveys, including in the time domain, are detailed in \citet{2013PASA...30...31B}.

In \S \ref{sec:surveystrategy} we describe the survey goals and observation strategy of GLEAM.
In \S \ref{sec:dataprocessing} we describe the image features and data processing issues associated with GLEAM.
In \S \ref{sec:sens} we calculate the thermal noise sensitivity of GLEAM mosaiced images for different observing frequencies, image weighting schemes and pointings.
In \S \ref{sec:outputs} we discuss outputs from GLEAM and compare to other radio surveys covering a large fraction of the southern hemisphere.

\section{Survey strategy}
\label{sec:surveystrategy}
An advantage for the survey speed of low-frequency radio telescopes is a large field-of-view.
For the MWA, whose main antenna `tiles' have an effective width of approximately 4 metres, this is especially true.
In addition, the 128 tiles of the MWA provide excellent snapshot $u,v$-coverage, which is essential for both the imaging and calibration fidelity of such wide-field data.
The combination of huge instantaneous field-of-view and excellent snapshot $u,v$-coverage makes the MWA well suited to surveying large volumes of the universe in a short time. 
As has been previously demonstrated \citep{2013ApJ...771..105B,2014MWACS}, meridian drift scans are an effective surveying technique for the MWA and we re-used the basic strategy for GLEAM.

\begin{figure*}
\hspace{-10mm}
\centering
    \begin{subfigure}[b]{0.33\textwidth}
                \label{fig:pb_69}
                \includegraphics[width=1.0\textwidth]{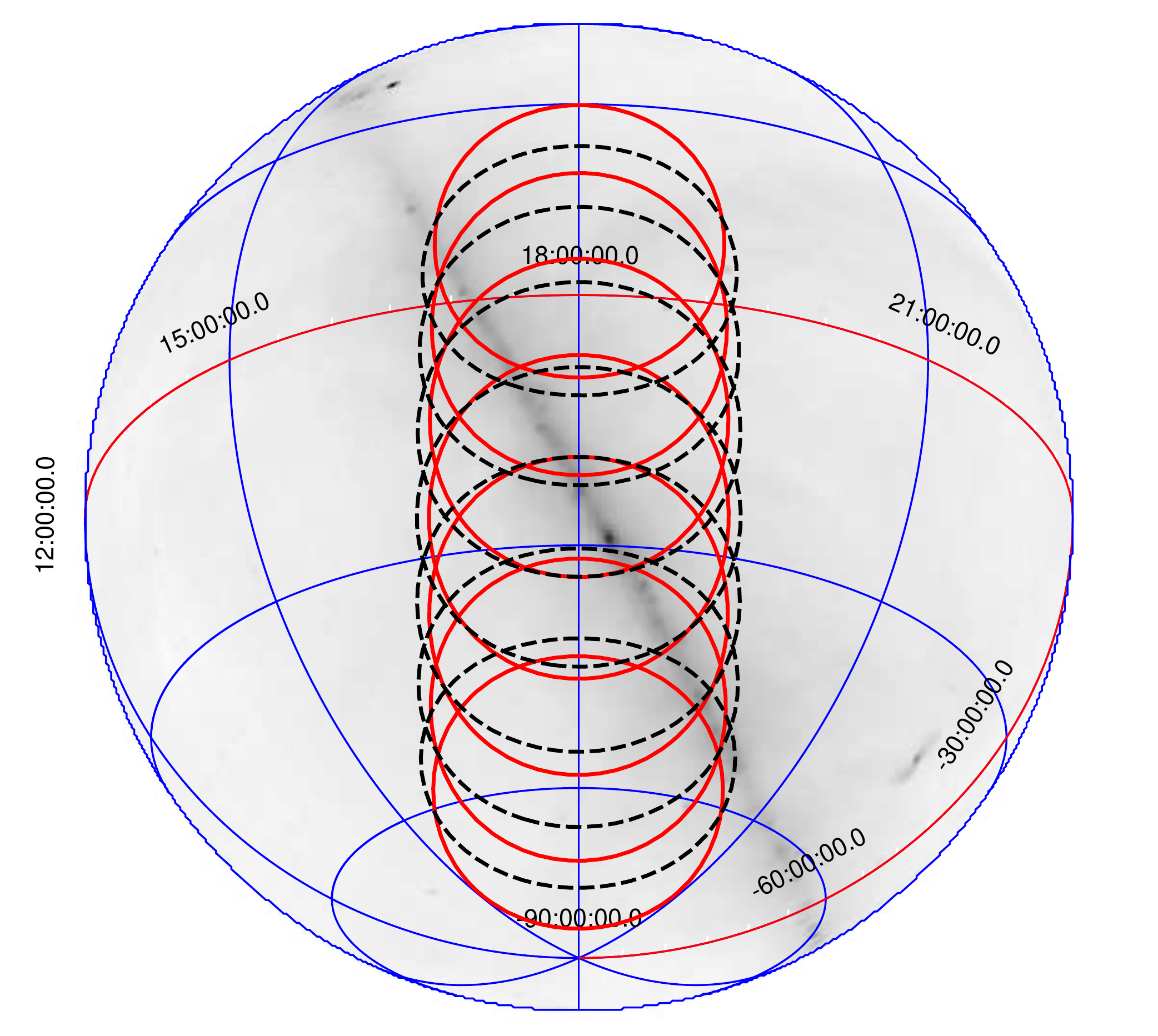}
		 \caption{88 MHz}
    \end{subfigure}
    \begin{subfigure}[b]{0.33\textwidth}
                \label{fig:pb_93}
                \includegraphics[width=1.0\textwidth]{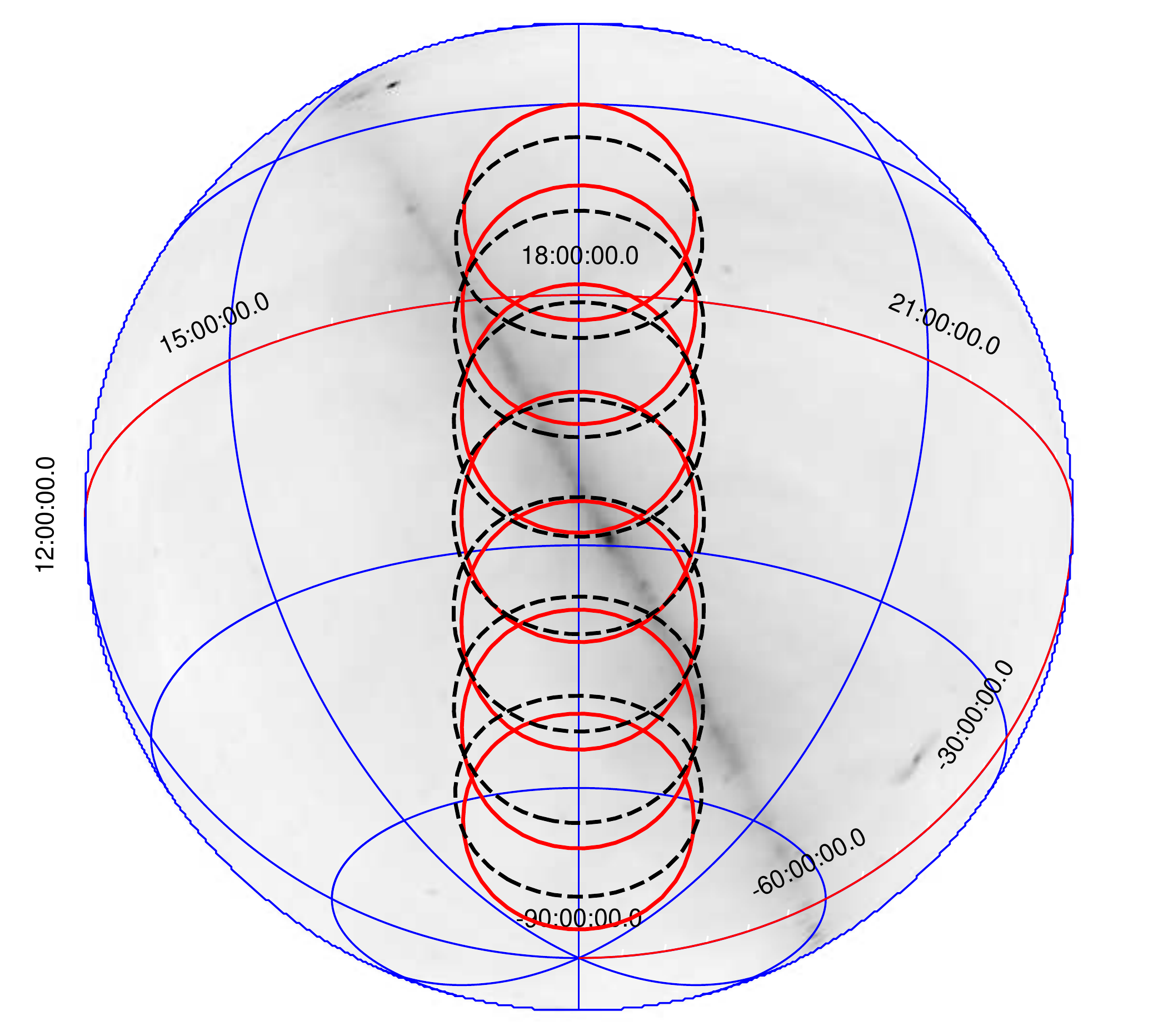}
		 \caption{118 MHz}
    \end{subfigure}
    \begin{subfigure}[b]{0.33\textwidth}
                \label{fig:pb_121}
                \includegraphics[width=1.0\textwidth]{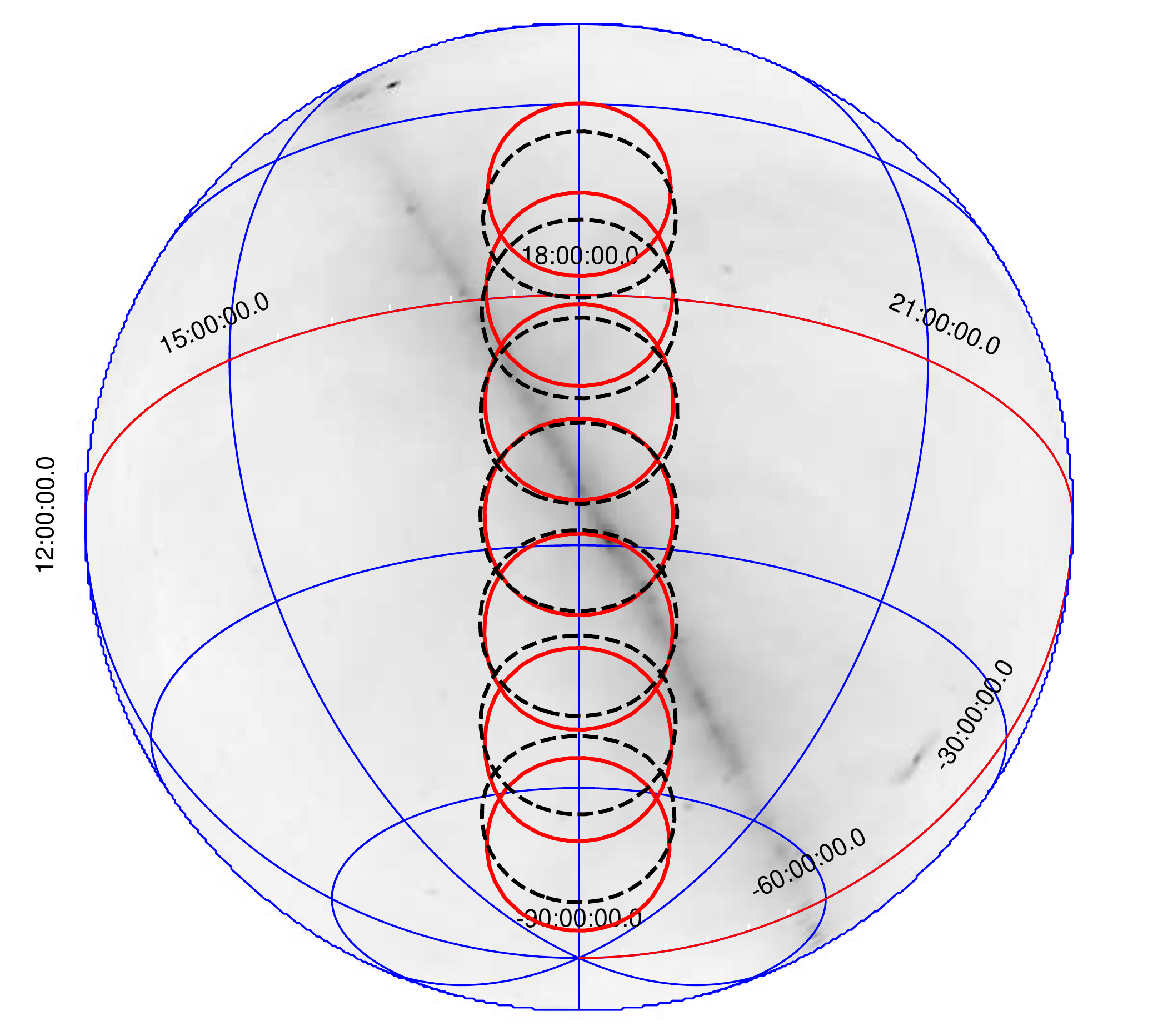}
		 \caption{154 MHz}
    \end{subfigure}

\hspace{-10mm}
    \begin{subfigure}[b]{0.33\textwidth}
                \label{fig:pb_145}
                \includegraphics[width=1.0\textwidth]{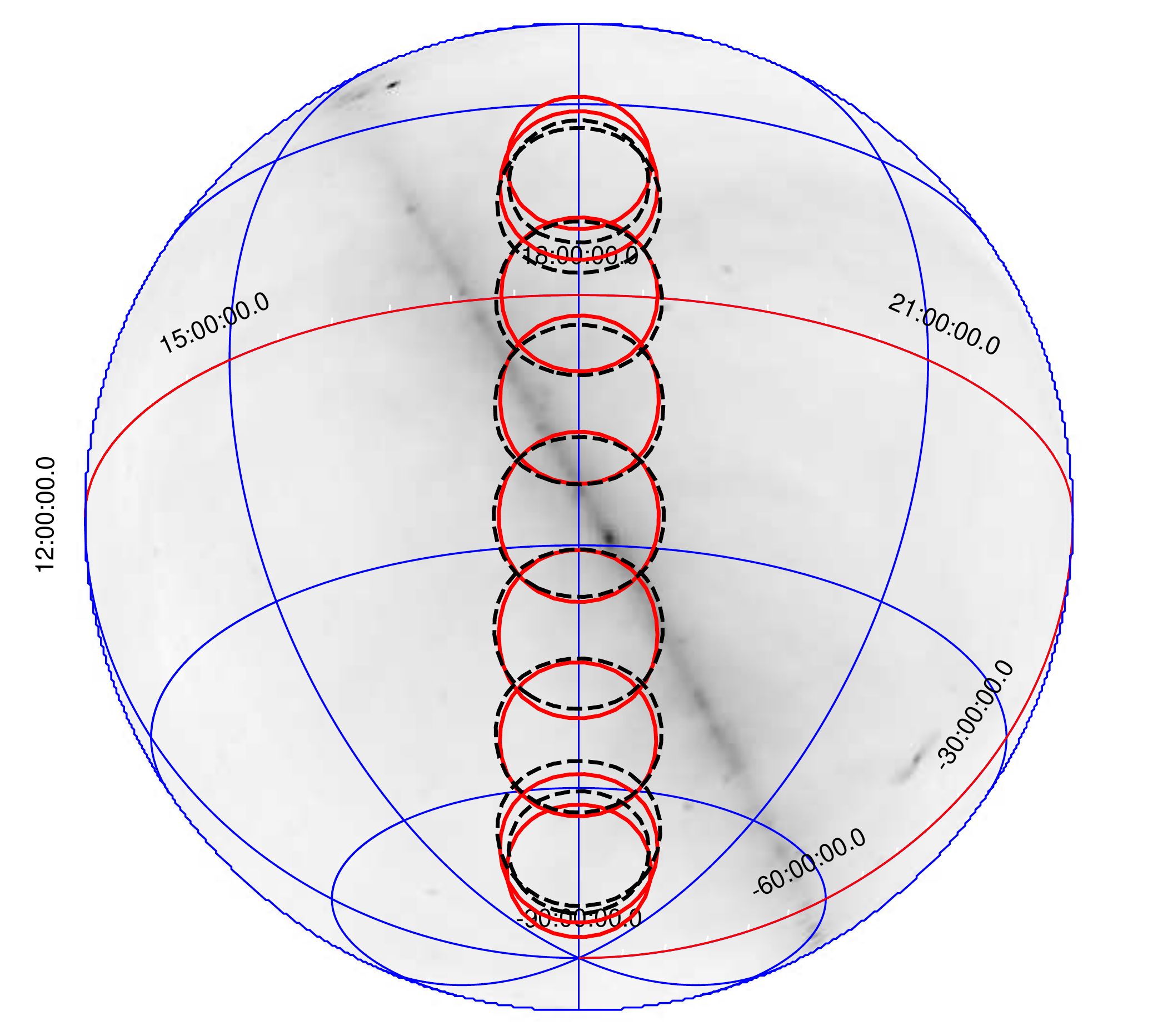}
		 \caption{185 MHz}
    \end{subfigure}
    \begin{subfigure}[b]{0.33\textwidth}
                \label{fig:pb_169}
                \includegraphics[width=1.0\textwidth]{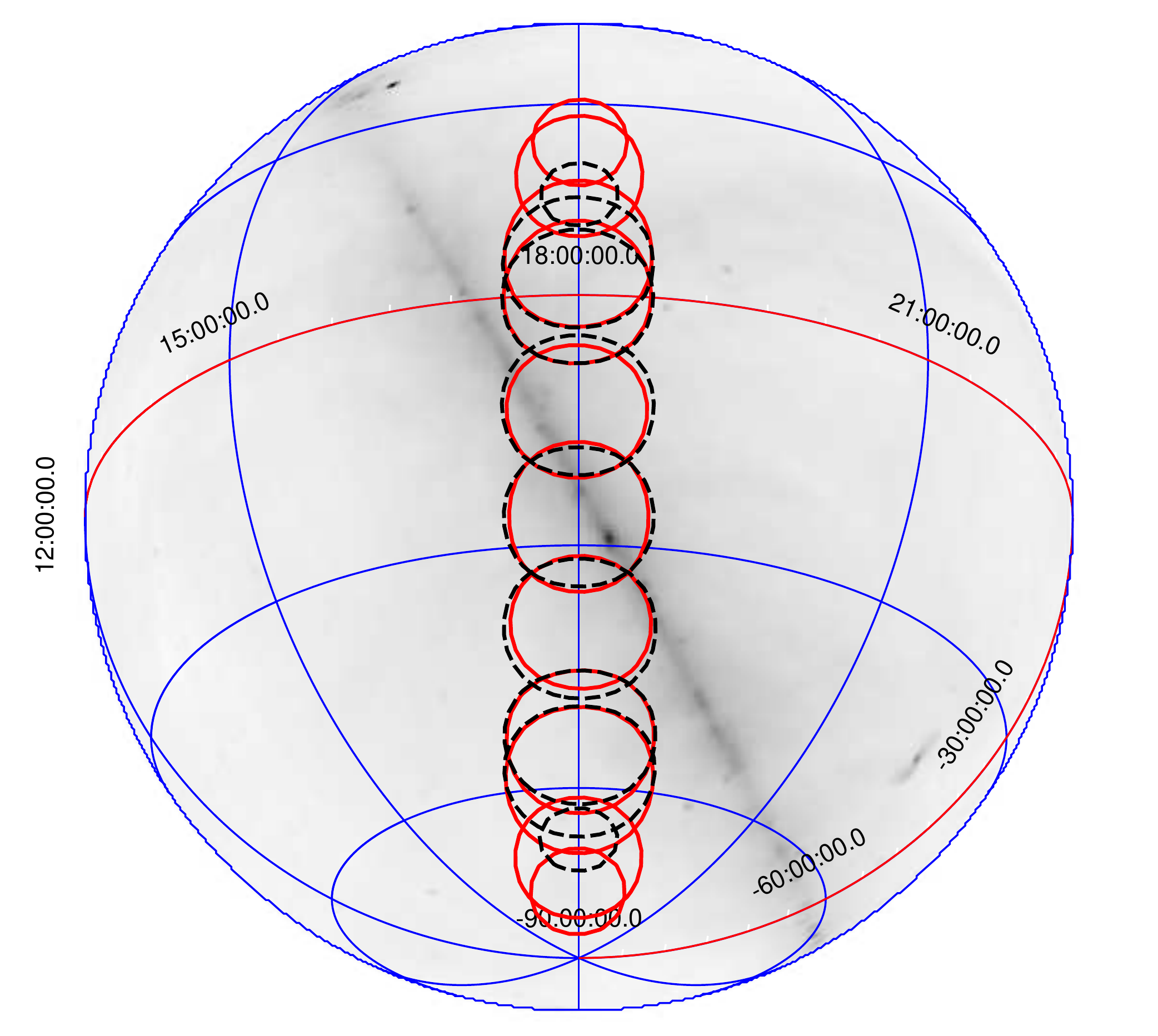}
		\caption{216 MHz}
    \end{subfigure}
    \begin{subfigure}[b]{0.33\textwidth}
                \label{fig:pb_Haslam}
                \includegraphics[width=1.0\textwidth]{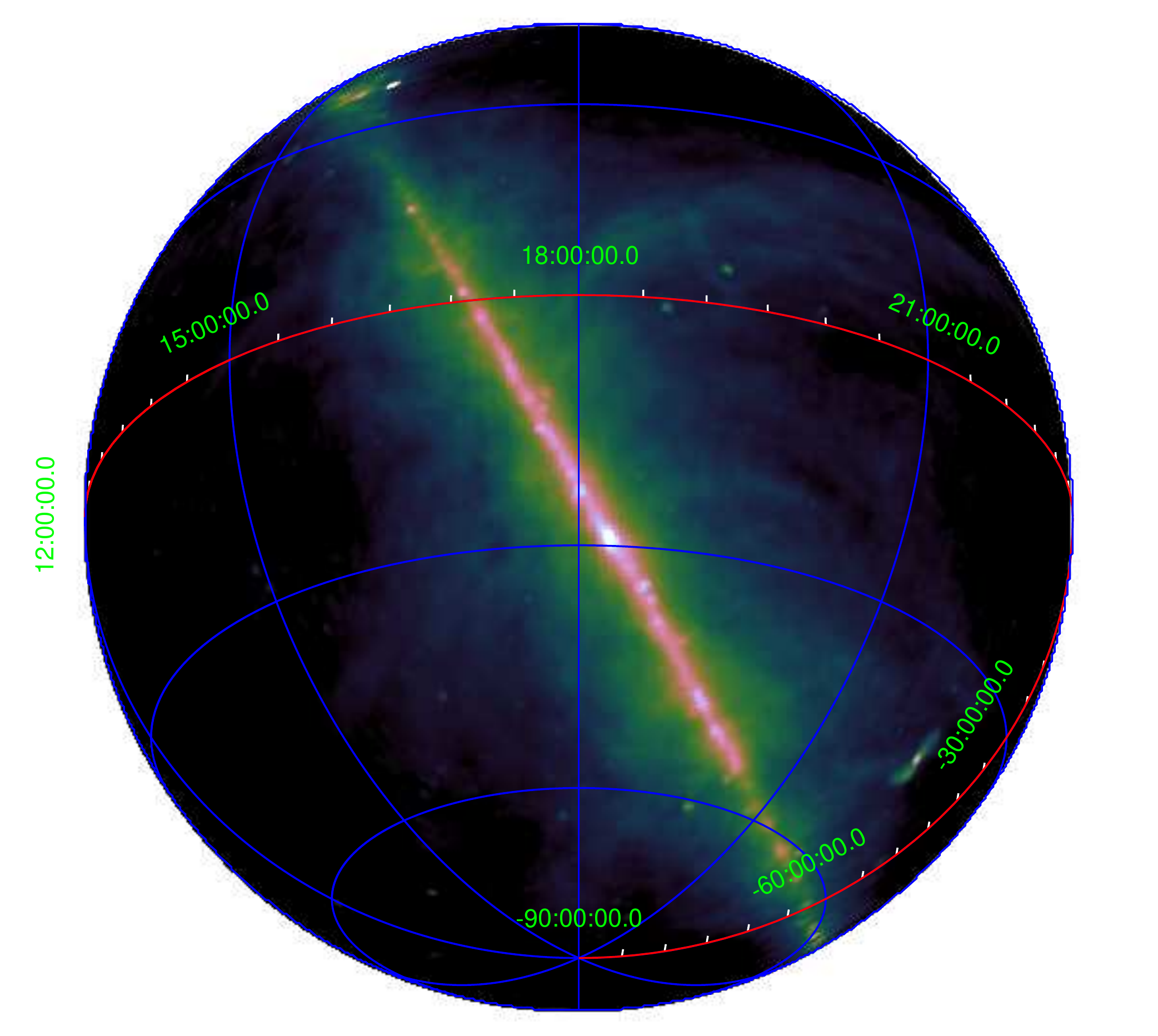}
		\caption{Example sky}
    \end{subfigure}
    \caption{Examples of the MWA's primary beams used in the survey overlaid on the sky at LST 18\,h as seen from the MWA site.
The contours show the half power levels relative to the peak in the beam for that declination where the solid (red) lines are for the `X' (east-west oriented) dipoles and the dashed black lines are for the `Y' (north-south oriented) dipoles.
The additional smaller ellipses at the extreme north and south for the higher frequency images are due to the chromatic grating lobes of the MWA primary beam exceeding the half power level of the main lobe.}
    \label{fig:primary_beams}
\end{figure*}

The sky was divided into seven strips in declination and five frequency ranges, as summarised in Table~\ref{tab:obs_summary}.
The declinations were chosen such that the peak in the primary beam response for a given setting corresponds approximately to the half power point of the neighbouring beam along the meridian at 150\,MHz, which also makes the beams overlap at the half power points at 216\,MHz. 
Examples of the primary beam patterns for the various frequencies and declinations are shown in Figure~\ref{fig:primary_beams}.

The instantaneous frequency coverage of the MWA is 30.72\,MHz, so the frequency range between 72 and 231\,MHz was divided into five bands that provide near contiguous coverage but avoid the band around 137\,MHz that is contaminated by satellite interference.
The central frequencies of the bands are listed in Table~\ref{tab:obs_summary}.

While GLEAM is designed to cover the entire sky south of $\delta$ $+25^{\circ}$, some sky north of $\delta$ $+25^{\circ}$ will be accessible with reduced sensitivity in the mid and lower parts of the GLEAM frequency range due to the larger primary beam size at those frequencies.

The region around the south celestial pole (SCP) is a special case for GLEAM due to its low elevation from the MWA site.
At the lower frequencies, the SCP region is included in the $-72^{\circ}$ declination observations.
For the two highest frequency ranges, extra observations pointed at the SCP are used to cover this region.

\begin{table*}
  \caption{GLEAM survey observing parameters.}
  \label{tab:obs_summary}
  \centering
  \begin{tabular}{ l | l} \hline \hline
    Pointing declinations (deg)  & -72, -55, -40.5, -26.7, -13, +1.6, +18.3 \\
    Central frequencies (MHz)    & 87.68, 118.4, 154.24, 184.96, 215.68 \\
    Frequency resolution (kHz)   & 40 (first year), 10 (second year) \\
    Time resolution (s)          & 0.5 (first year), 2 (second year) \\
  \hline \hline
  \end{tabular}
\end{table*}

GLEAM observing was executed as a series of week-long campaigns where a single declination setting was observed in a night, covering a strip between approximately 8 and 10 hours in length, depending on the time of year.
Radio emission from the Sun can overwhelm other radio sources at MWA frequencies, so observations were only performed at night.
Within a night, the observing was broken into a series of 120\,s scans for each frequency, cycling through all five frequency settings over 10 minutes.
Within a scan, typically 108\,s of usable data were collected.
Every two hours throughout the night, a calibration field containing a bright, compact source was observed over all five frequency settings, again as a set of 120\,s scans totalling 10 minutes.

During the first year of GLEAM observations, visibility data were recorded with the default MWA time and frequency resolution of 0.5\,s and 40\,kHz respectively.
This covered the entire sky as originally planned for GLEAM.
In the second year of GLEAM, enhancements to the MWA correlator \citep{2015PASA...32....6O} allowed data to be recorded in 10\,kHz mode, hence the second year of GLEAM was recorded with 2\,s time resolution and 10\,kHz frequency resolution, increasing the usefulness of the dataset for spectral line and polarisation science.
In addition, for the second year of GLEAM the antennas were pointed off meridian by $+1$ or $-1$ hours in alternating observing campaigns to increase the overall hour angle coverage over the entire sky.

GLEAM observations were also used commensally as part of the MWA Transients Survey (`MWATS', Bell et al., in prep), which is a time domain survey combining GLEAM data with separate observations to regularly monitor the radio sky.
The aim of MWATS is to revisit three of the GLEAM declination strips ($+1.6^{\circ}$, $-26.7^{\circ}$ and $-55^{\circ}$) on timescales of seconds, hours and months using data centred at 154\,MHz only.
The motivations for this survey are as follows: (i) To obtain temporal data on an extremely large and robust sample of low-frequency sources to explore and quantify both intrinsic and extrinsic variability; (ii) To search and find new classes of low-frequency radio transients that previously remained undetected and obscured from multi-wavelength discovery; (iii) To place rigorous limits on the occurrence of both transients and variables prior to the SKA era.

\section{Data processing and mosaicing}
\label{sec:dataprocessing}

\begin{figure*}
    \begin{subfigure}[b]{0.5\textwidth}
                \label{fig:uv_mono}
	\hspace{-5mm}
                \includegraphics[width=1.0\textwidth]{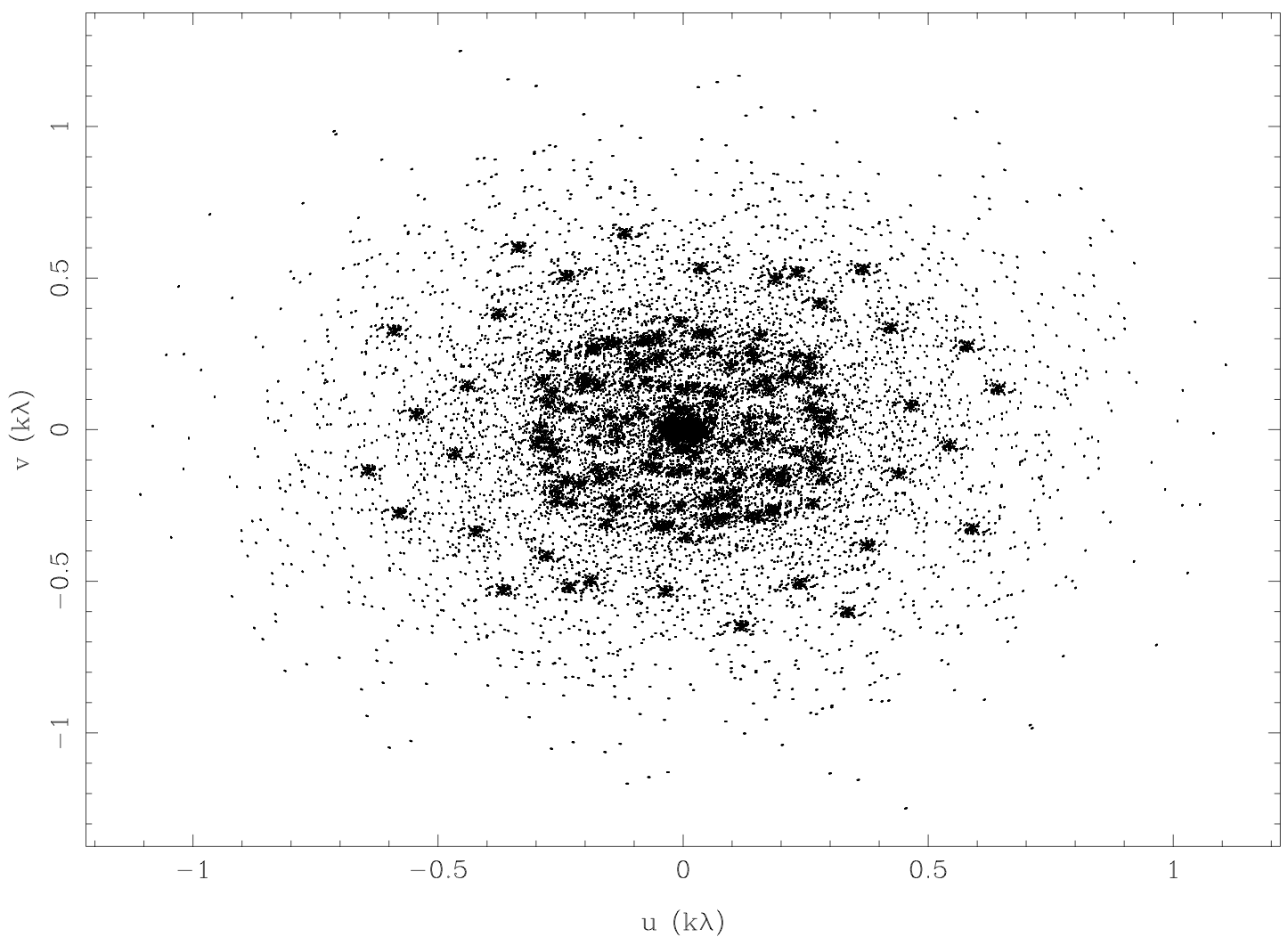}
    \end{subfigure}
    \begin{subfigure}[b]{0.5\textwidth}
                \label{fig:uv_mfs}
	\hspace{-1mm}
                \includegraphics[width=1.0\textwidth]{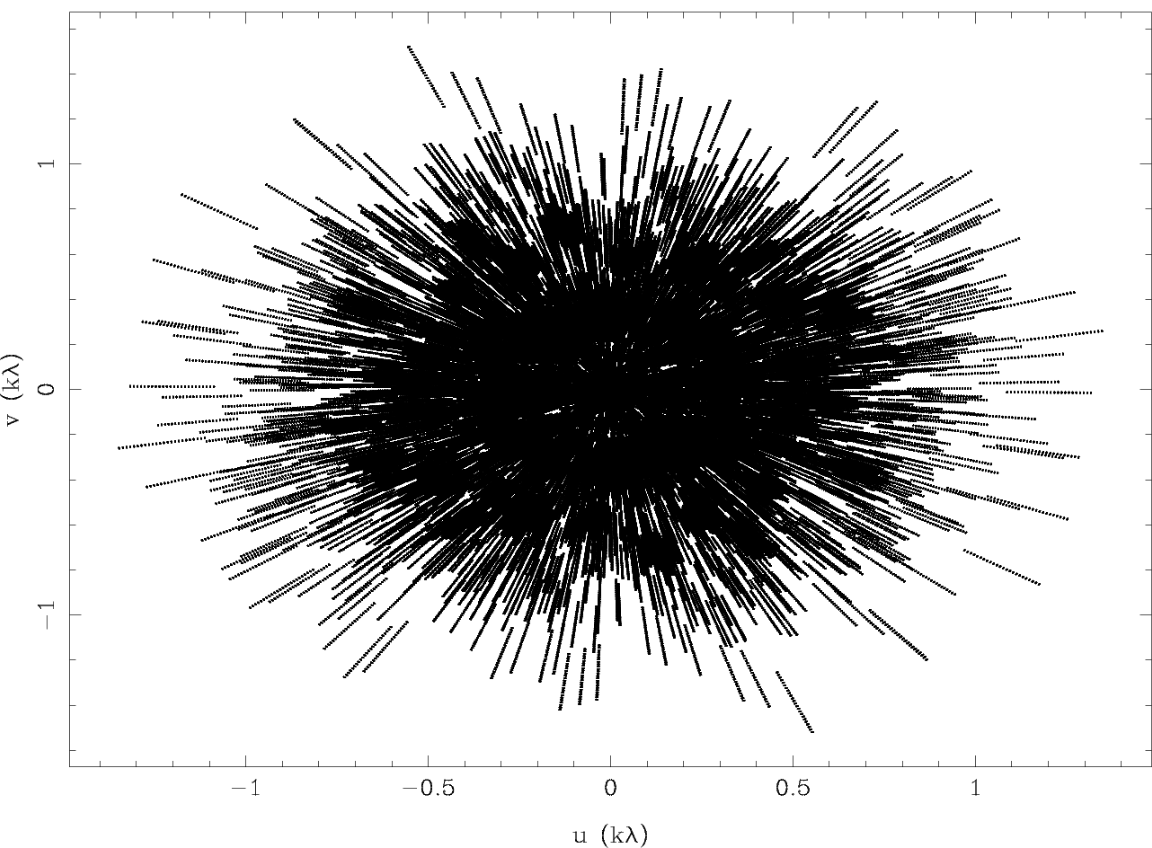}
    \end{subfigure}
    \caption{MWA monochromatic (left) and 30\,MHz multi-frequency synthesis (right) $u,v$-coverage for a 2\,minute zenith pointed snapshot centred on 154\,MHz.}
    \label{fig:uv_cov}
\end{figure*}

Here we briefly review the main features of the MWA and refer the reader to \citet{2013PASA...30....7T} for details.\\
The MWA's 128 antenna tiles are distributed over an area approximately 2.5\,km in diameter.
The array has a central core of approximately 40\% of the tiles with the remainder distributed for $u,v$-coverage.
An example of $u,v$-coverage for a 154\,MHz zenith pointed snapshot is shown in Figure~\ref{fig:uv_cov}.
The excellent instantaneous $u,v$-coverage of the MWA is advantageous for both calibration and imaging.

As discussed in \citet{2010PASP..122.1353O}, MWA snapshot images follow the well-defined slant orthographic projection which is supported by most current astronomy software.
The simplest approach for mosaicing GLEAM images is image-based weighted addition, after regridding to a common coordinate frame.
This approach was used in \citet{2014MWACS} and we adopt the same approach here to calculate sensitivity (see \S\ref{sec:sens}).

Data are typically pre-processed through the \textsc{Cotter} pipeline \citep{2015PASA...32....8O} which flags radio-frequency interference (RFI) and optionally reformats the data into standard radio astronomy data formats.

\begin{figure*}
    \begin{subfigure}[b]{0.5\textwidth}
                \label{fig:zen_snap_full}
    \hspace{-7mm}
                \includegraphics[width=\textwidth,scale=1.01]{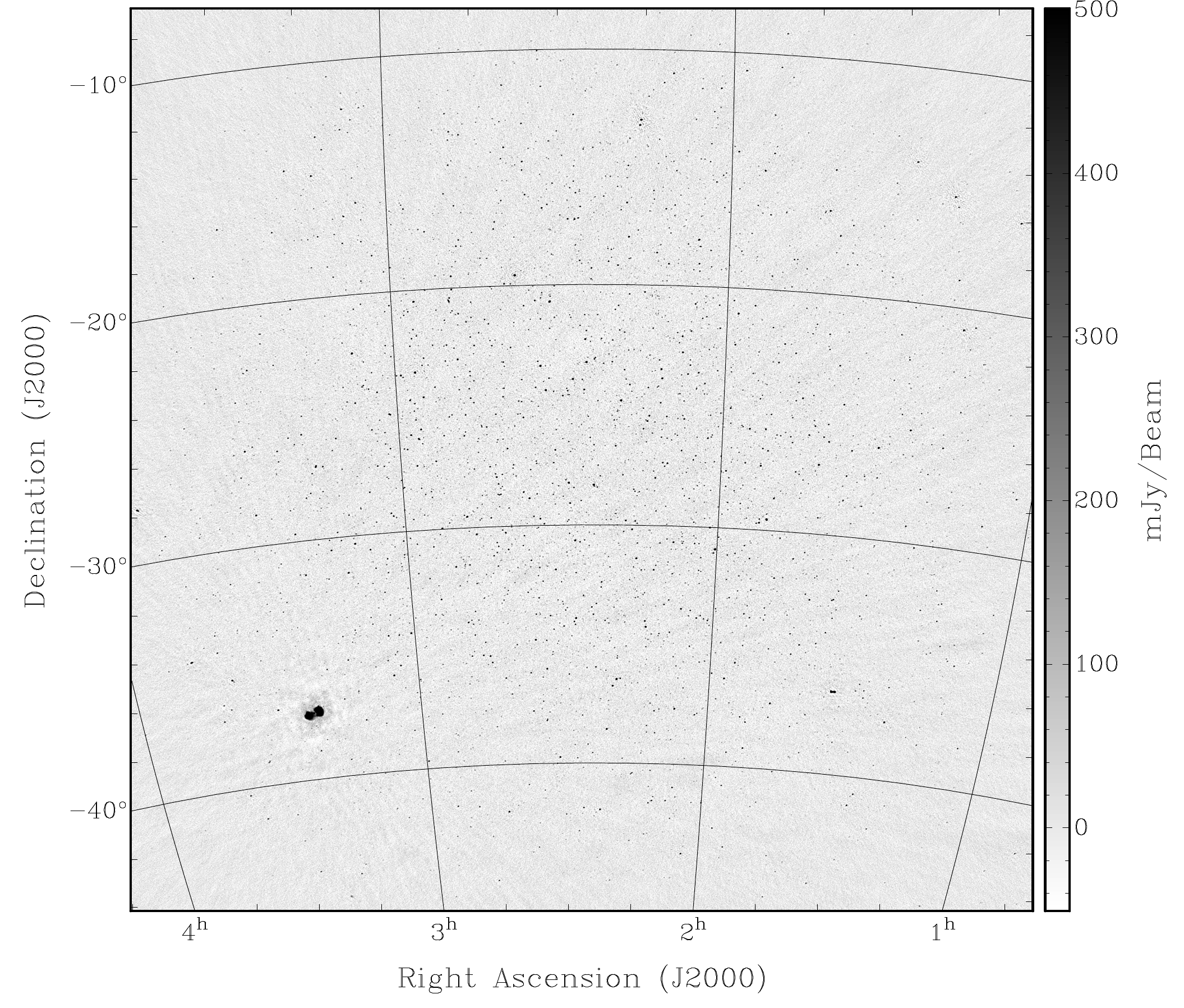}
    \end{subfigure}
    \begin{subfigure}[b]{0.5\textwidth}
                \label{fig:zen_snap_zoom}
                \includegraphics[width=\textwidth]{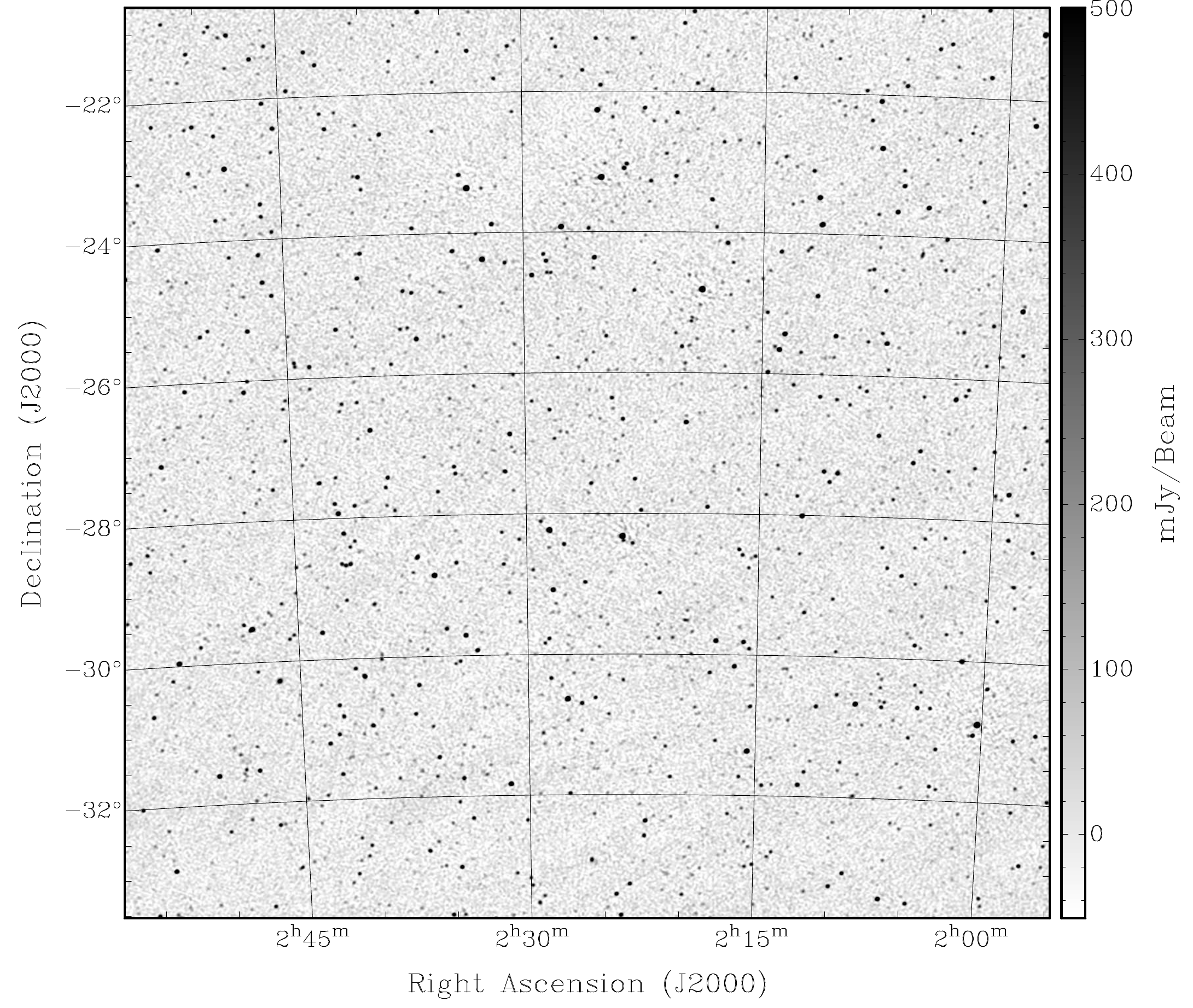}
    \end{subfigure}
    \caption{An example of a high Galactic latitude zenith-pointed 2\,minute snapshot for a single (YY) polarisation centred at 154\,MHz using full bandwidth multi-frequency synthesis with robust parameter $= -1$. The left panel shows the full $4096 \times 4096$ image (not primary beam corrected), which extends almost to the first beam null. The right image is zoomed to the field centre. The beam size in these images is $2.46 \times 2.24$ arcmin.}
    \label{fig:zen_snap}
\end{figure*}

Fig~\ref{fig:zen_snap} shows example snapshot images from GLEAM at high Galactic latitude.
Images containing almost the entire main lobe of the primary beam can be generated with standard ($\sim \times3$) oversampling of the synthesised beam in image space and $4096\times4096$ image pixels.
The beam size in Fig~\ref{fig:zen_snap} is $2.46 \times 2.24$~arcmin and the pixels are 34 arcsec across at the image centre.
In this example, the giant radio galaxy Fornax A can be seen in the lower left of the image, however the vast majority of extragalactic radio sources are unresolved in GLEAM.

\begin{figure*}
    \begin{subfigure}[b]{0.5\textwidth}
                \label{fig:Vega_snap_nozero}
		\hspace{-8mm}
                \includegraphics[width=0.875\linewidth,angle=-90]{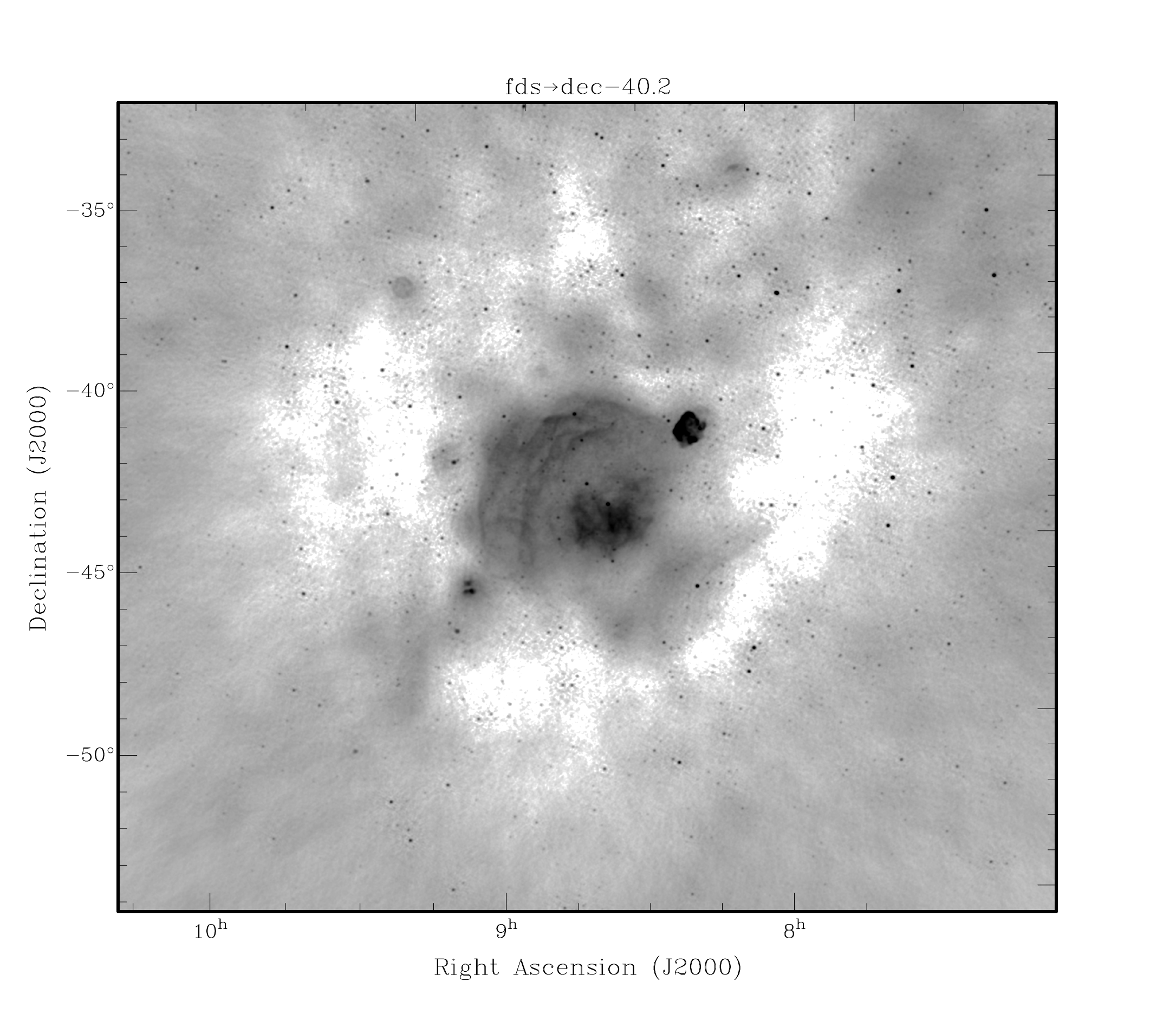}
    \end{subfigure}
    \begin{subfigure}[b]{0.5\textwidth}
                \label{fig:Vela_snap_withzero}
		\hspace{-3mm}
                \includegraphics[width=0.875\linewidth,angle=-90]{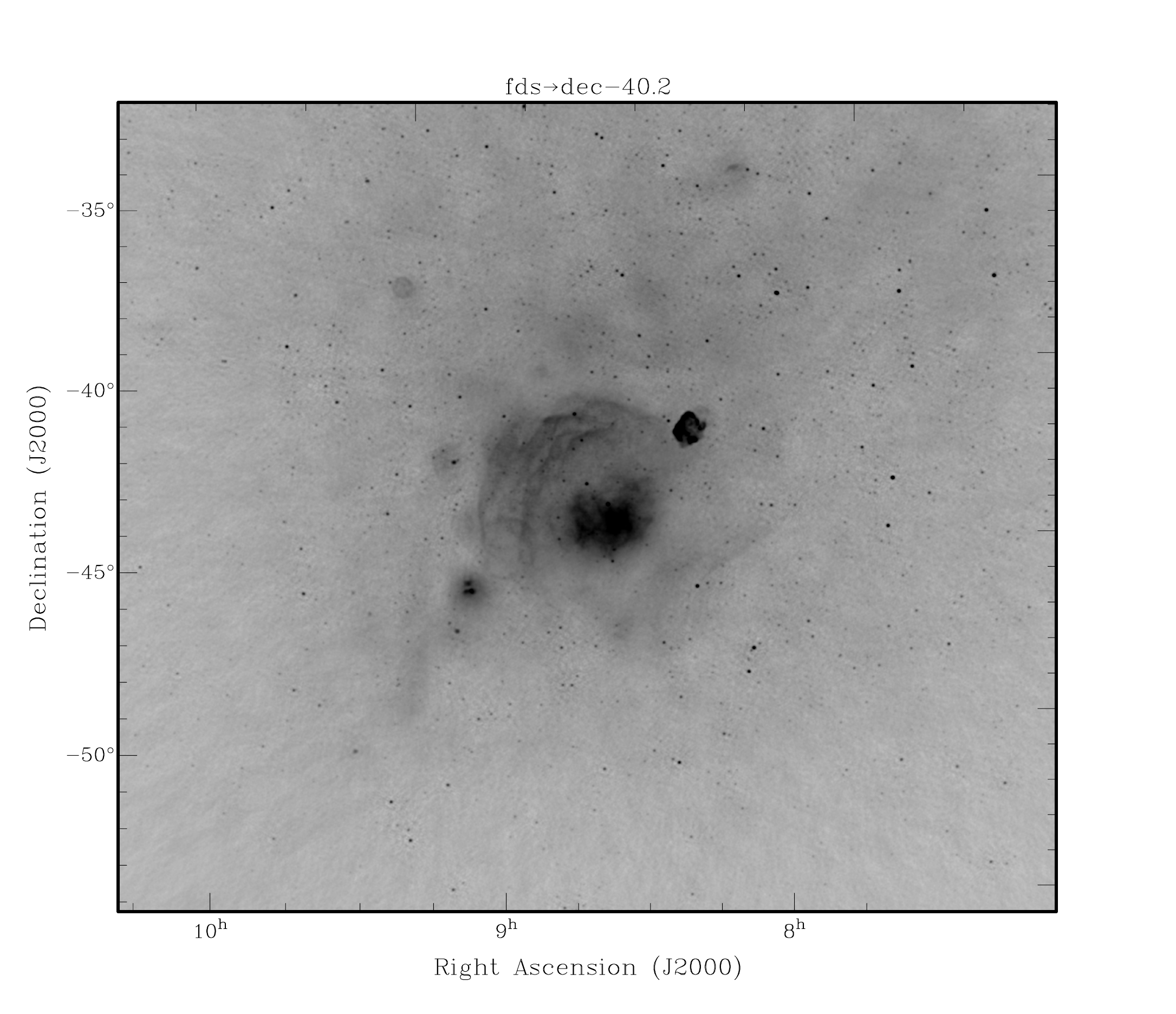}
    \end{subfigure}
    \caption{An example of the Gum Nebula region of the Galactic plane including the Vela and Puppis~A supernova remnants from a 2\,minute snapshot in the declination $-40^{\circ}$ drift scan centred on 154\,MHz. Images were made with full bandwidth synthesis over 30.72\,MHz and robust\,$=0.5$ weighting.
The left image shows a restored image after deconvolution with no additional zero-spacing information.
The right image shows the result after using \textsc{Miriad}'s `immerge' task to include zero-spacing information based on 408\,MHz data \citep{1982A&AS...47....1H}.
The beam size in these images is $4.22 \times 3.97$ arcmin. The colour bar has been omitted because the absolute flux scale is not yet correct, but the two images have the same intensity scaling.}
    \label{fig:Vela_snap}
\end{figure*}

Figure~\ref{fig:Vela_snap} shows example snapshot images from GLEAM on the Galactic plane region containing the Gum Nebula from the declination $-40^{\circ}$ drift scan.
The left image is the restored image just using MWA data and shows a negative bowl on the largest scales around the region due to missing zero-spacing flux.
The right image shows the same region with the addition of zero-spacing information using \textsc{Miriad}'s `immerge' task \citep{1995ASPC...77..433S} using a scaled, regridded and primary beam weighted image from 408\,MHz data \citet{1982A&AS...47....1H}.
The 408\,MHz data were scaled by a constant spectral index of -0.6, however since Figure~\ref{fig:Vela_snap} is only meant for illustrative purposes we allowed `immerge' to adjust the flux scales of the datasets in an overlap region for baseline lengths between 10 and 45 meters, hence the overall flux scale is only approximate.
These images demonstrate the MWA's excellent surface brightness sensitivity.


\subsection{Calibration}
\label{sec:calibration}

The GLEAM observation strategy includes a calibration scan on a bright compact source every two hours.
The MWA is phase stable over many hours and phase calibration solutions are readily transferred between pointings, after which self-calibration can reduce residual phase and amplitude calibration errors.
Calibration sources used for GLEAM include 3C444, Pictor A, Taurus A, Hydra A, Virgo A and Hercules A, most of which have well determined flux densities over the GLEAM frequency range.
These scans are useful for diagnostic purposes and are used to apply an initial amplitude and phase calibration solution onto the survey data.

Since the absolute response of the MWA's primary beam changes depending on pointing, the flux scale of snapshots made via calibration transfer must be scaled by the ratio of the absolute primary beam response on the target field to the absolute primary beam response on the calibration field.
This procedure was used to set the flux scale of the example snapshot images used in this paper but is limited by how accurately the model primary beam represents the true primary beam.

The full survey, however, has the advantage that many thousands of sources sample the primary beam response as they drift through it during the night and that the primary beam response is constant throughout the night.
This provides a mechanism to build an accurate empirical model of the primary beam and simultaneously bootstrap the absolute flux scale using a large sample of sources.
The overlapping declination ranges of GLEAM also provide a way to ensure both the primary beam correction and overall flux scale is consistent over the entire sky.

Some complications do arise in processing GLEAM snapshots into mosaics, in particular: effects of the ionosphere; strong sources in the primary beam sidelobes; and instrumental polarisation.

Concerning the ionosphere, the MWA is located at the Murchison Radio-astronomy Observatory (MRO), the location of which was chosen in part due to its favourable ionospheric characteristics\footnote{\url{https://www.skatelescope.org/wp-content/uploads/2012/06/64_Appendix-2.5.1.pdf}} in the mid-latitude southern hemisphere.
For GLEAM purposes, the ionosphere over the MRO is typically stable \citep{2014HerneConf} with slowly varying changes in Total Electron Content (TEC) causing arcmin-scale shifts in source positions over timescales of hours \citep{2014MWACS,2015Singh}, which is of order the size of the MWA synthesised beam.
Under typical conditions these shifts can be described as a single bulk position offset for all sources in the field, an effect which can be corrected for in an image simply by adjusting the reference coordinates in the image metadata.
Unusual ionospheric, tropospheric, and solar wind conditions do occasionally occur and are the subject of specific detailed studies \citep{2015arXiv150406470T}.
Nights that showed detrimental ionospheric behaviour were re-observed.

Powerful radio sources (Cygnus A, Centaurus A, Virgo A, Taurus A etc) in the MWA's sidelobes can cause artifacts in multi-frequency synthesis images that cannot be deconvolved.
These artifacts are due to chromatic effects and the nature of the MWA's primary beams where, after calibration of the main lobe of the beam, differences between antenna tiles manifest themselves as differences in the sidelobes of the primary beam.
These bright sources can be modelled and subtracted, in principle, via advanced interferometric techniques such as `peeling' and `A-projection' \citep{2007ITSP...55.4497V,2008ISTSP...2..707M,2008A&A...487..419B,2013A&A...553A.105T}, but are a problem for conventional deconvolution where all primary beams are assumed to be identical.
For GLEAM, such sources are a problem in a fraction of data ($\sim 25\%$).

\subsection{Polarisation}
\label{sec:gleampol}

Since the MWA's antenna tiles are fixed on the ground, the MWA's primary beam is subject to many significant instrumental polarisation effects.
Full polarisation data products in instrumental coordinates (XX,YY,XY and YX) are stored by default from the MWA correlator \citep{2015PASA...32....6O}. 
The two main issues for polarisation are instrumental cross polarisation, due mostly to geometric effects, and a difference in the magnitude of the co-polarisation (XX and YY) response within a tile.
The cross polarisation is an inevitable projection effect which causes an unpolarised source to generate correlated signal in the cross-polarisation correlator outputs (XY and YX) independent of any electronic leakage.
The difference in co-polar response (XX and YY) is expected partly due to simple geometric effects (i.e. a dipole's effective length shortens as a source moves closer to its long axis) and array mutual coupling effects.
Both of these effects are considered in the tile model detailed in \citet{2015RaSc...50...52S}, and this model has been adopted as the standard for all MWA data processing.

Processing with appropriate calibration and imaging techniques \citep[e.g.][]{2008ISTSP...2..707M,2014MNRAS.444..606O}, to account for direction-dependent and time-dependent effects, allows both linear (Stokes Q, U) and circular polarisation (Stokes V) images to be produced.
As the polarised source counts are significantly lower than in total intensity (Stokes I), linear and circular polarisation images are in principle far less affected by source confusion and can achieve sensitivity levels that approach thermal noise.

Circular polarisation can easily be processed in a similar manner to that used for total intensity (Stokes I) imaging, that is, by treating the entire frequency band in a continuum-like mode. However, care must be taken when imaging the two linear polarisation (Stokes Q and U) as even small degrees of Faraday rotation can significantly rotate the polarisation angle of the radiation across the observing bandwidth, resulting in significant bandwidth depolarisation.
To reduce the effect of bandwidth depolarisation it is necessary to image Q and U in a spectral-like mode using the fine (10 or 40\,kHz) channels available in the GLEAM data. Rotation measure (RM) synthesis \citep{2005A&A...441.1217B} may then be used to recover the linear polarisation maps from the resulting RM cube at selected Faraday depths ($\phi$), as was previously demonstrated by \citet{2013ApJ...771..105B}.

\begin{figure*}
\hspace{-10mm}
\includegraphics[width=1.1\linewidth]{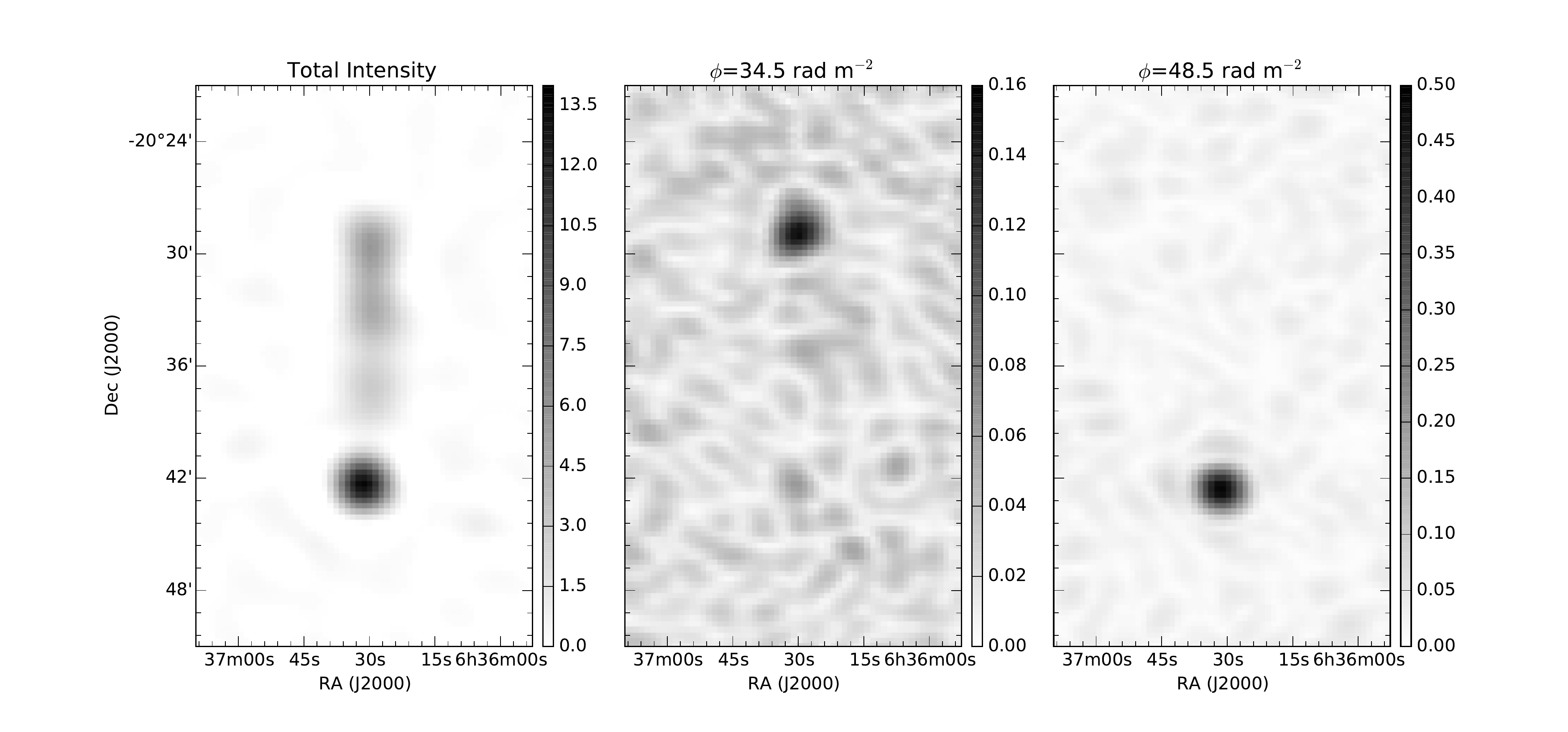}
\caption{Total intensity map of PMN J0636-2041 and the associated polarised intensity maps of the source taken at two different Faraday depths
($\phi$=34.5\,rad\,m$^{-2}$ and 48.5\,rad\,m$^{-2}$). All images were processed using uniform visibility weighting in the $138.88-169.60$\,MHz band.
Units are Jy\,beam$^{-1}$ for total intensity and Jy\,beam$^{-1}$\,RMSF$^{-1}$ for polarised intensity.}
\label{fig:polim}
\end{figure*}

Figure~\ref{fig:polim} presents an example polarised source, PMN\,J0636-2041, detected with early GLEAM data. The total intensity image was processed using continuum-mode imaging and shows a source with an extended morphology. The linear polarisations were processed using spectral mode imaging at 40 kHz spectral resolution and RM synthesis.
The polarised intensity images, selected from the resulting RM cube, show that the northern and southern components of this source exhibit peaks at two different Faraday depths ($\phi=34.5\pm0.1$\,rad\,m$^{-2}$ and $\phi=48.5\pm0.05$\,rad\,m$^{-2}$ respectively.)
These are both consistent with VLA observations of this source at 1.4\,GHz \citep{2009ApJ...702.1230T}.

The resolution ($\delta\phi$), maximum scale size sensitivity (max. scale) and Faraday depth range ($\lVert\phi_{max}\rVert$) available when using the RM synthesis technique is a function of the channel width, the bandwidth and the highest frequency channel available \citep{2005A&A...441.1217B}. This has been summarised in Table \ref{table:polpar} for each of the GLEAM observing bands. Note that maximum scale size is always smaller than the resolution in $\phi$ space. Essentially, this means that the MWA cannot resolve Faraday thick clouds in $\phi$ space in any single band. Where sensitivity to wider structures in Faraday space is required, lower frequency bands can be combined with the upper frequency band. In this instance, the maximum scale size is set by the upper band (1.9\,rad\,m$^{-2}$) but the resolution is increased by the separation of the bands.

While high precision is achievable in Faraday space it should be noted that the true Faraday depth measured will be affected by the ionosphere. The magnitude of the ionospheric degradation can be reduced by observing at night and at zenith. However, applications that require high-accuracy measurements of Faraday depth will need to correct for the ionospheric component.
Observations with the 32-tile commissioning array confirmed that codes that predict the ionospheric Faraday rotation, e.g. ionFR \citep{2013A&A...552A..58S} and ALBUS\footnote{\url{https://github.com/twillis449/ALBUS_ionosphere}}, are consistent with the observed effect on linearly polarised sources.
During the course of a GLEAM observation, the ionospheric component of the Faraday rotation does not vary significantly; typical variations are of order $\sim0.3$\,rad\,m$^{-2}$.
However, the overall ionospheric Faraday rotation component can vary significantly from epoch to epoch, often by several rad\,m$^{-2}$, and should be corrected for.

All polarisation observations are affected by some level of polarisation leakage from one Stokes parameter to another.
For dipole instruments such as the MWA, polarisation leakage is a frequency-dependent and position-dependent effect resulting from errors in the primary beam model used for calibration.
The effect results in a proportion of the Stokes I signal ``leaking'' into Stokes, Q, U and V.
In Faraday space this results in an increased flux density at $\phi=0$\,rad\,m$^{-2}$.
The magnitude of leakage increases with frequency and with angular distance from zenith.
In the $139-170$\,MHz band, the leakage is of order $1-2\%$ within $10^{\circ}$ of zenith but can increase to $\sim5\%$ at the survey extremities \citep{2015RaSc...50...52S}.

The GLEAM observations provide an outstanding data set for exploring diffuse polarised emission from the Milky Way.
Even on a per-snapshot basis, the densely sampled core of the full 128-tile MWA array has excellent sensitivity to large-scale structures that may be expected from diffuse galactic polarised emission.
\cite{2013ApJ...771..105B} observed such emission with the 32-tile MWA prototype and found the total polarisation surface brightness peaking at $\sim200$\,mJy\,beam$^{-1}$.
By employing a natural weighting scheme to improve sensitivity to large-scale structures, the MWA array can image such features with a signal to noise ratio of $\sim70$ in a single 2-minute snapshot. This level of sensitivity may indeed be sufficient to observe variations in the ionosphere by observing variations in Faraday rotation in the diffuse Galactic background.

\begin{table}
\centering
\caption{Polarisation parameters for GLEAM observing bands with 40\,kHz frequency resolution.}
\begin{tabular}{r c c c c}
\hline\hline
GLEAM Band & $\delta\phi$ & max. scale & $\lVert\phi_{max}\rVert$ \\ [0.5ex]
(MHz)      & (rad\,m$^{-2}$) & (rad\,m$^{-2}$) & (rad\,m$^{-2}$) \\ [0.5ex]
\hline
$72.30-103.04$  & 0.40 & 0.37 & 91.1 \\
$103.04-133.76$ & 1.0 & 0.63 & 263.7 \\
$138.88-169.60$ & 2.3 & 1.0 & 645.6 \\
$169.60-200.32$ & 3.9 & 1.4  & 1175.6 \\
$200.32-231.04$ & 6.2 & 1.9  & 1937.0 \\ [1ex]
\hline\hline
\end{tabular}
\label{table:polpar}
\end{table}

\subsection{Comparison of the GLEAM and ATLAS images in the CDFS}

The third data release from the Australia Telescope Large Area Survey \citep[ATLAS DR3;][]{2015Franzen} covers
an area of 3.6\,$\mathrm{deg}^{2}$ coincident with the Chandra Deep Field South 
(CDFS; $\alpha = 03^{\rm h} 30^{\rm m} 16.3^{\rm s}$; $\delta = -28^{\rm o} 05' 12.4''$ J2000) to a typical sensitivity
of $14~\mu \mathrm{Jy}$\,$\mathrm{beam}^{-1}$ at 1.4\,GHz.
We have compared the ATLAS image of the CDFS with the declination $-27^{\circ}$ GLEAM mosaic at $147-154$\,MHz,
which is closely matched in declination to the CDFS.
Given the much higher sensitivity of ATLAS, we expect all GLEAM sources to be detected in ATLAS.
We compared source positions in the two images but not source flux densities since the absolute flux density scale for GLEAM is only approximately correct.

\begin{figure*}
\begin{center}
\includegraphics[scale=0.95,angle=270]{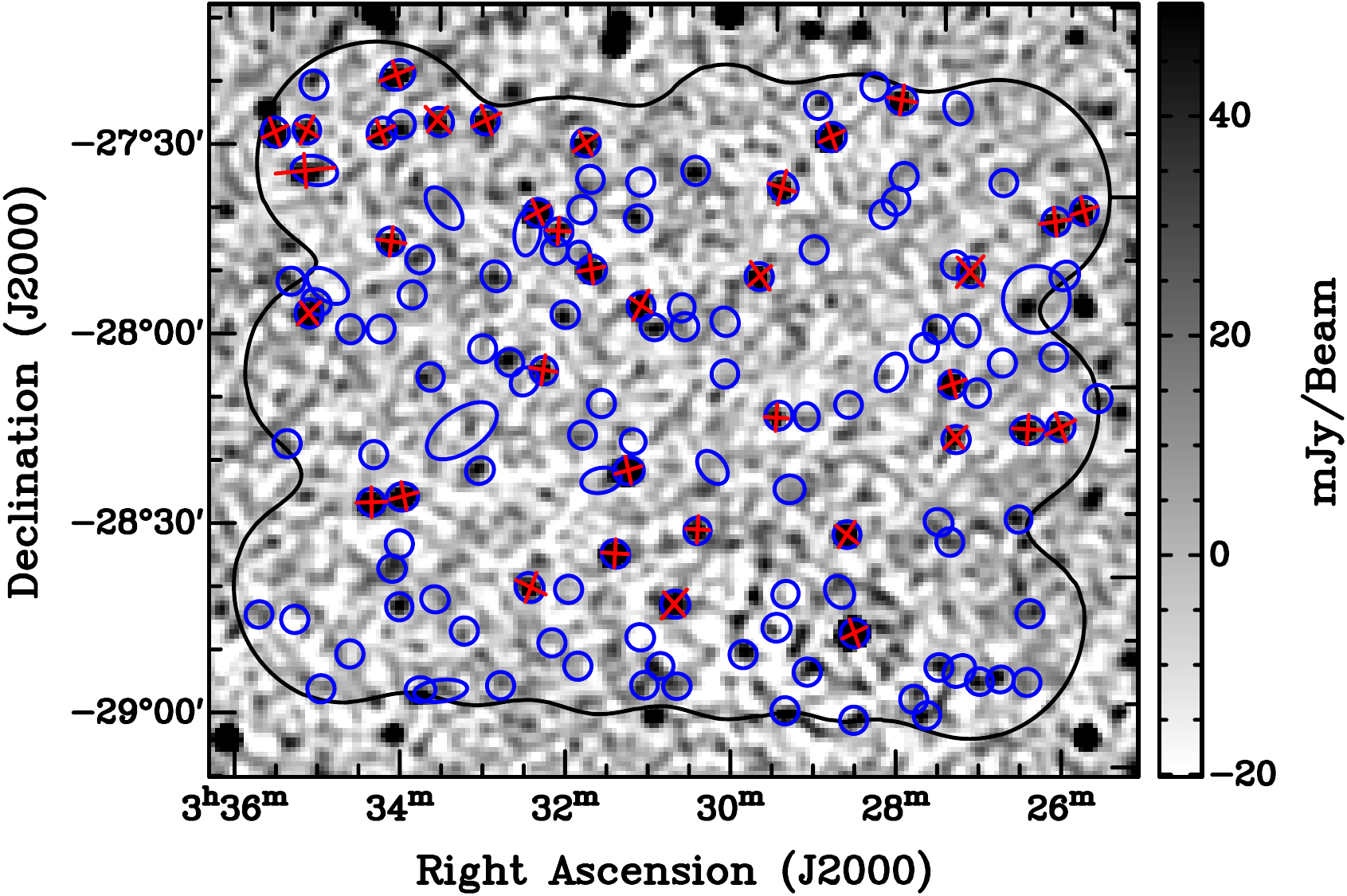}
\end{center}
\caption {Section of a $147-154$~MHz declination $-27^{\circ}$ XX mosaic coincident with the CDFS.
The greyscale is linear and runs from $-20$ to $+50$\,mJy (the overall flux scale is only approximately correct).
GLEAM detections above $5 \sigma$ are shown as red crosses of the same shapes as their fitted Gaussian 
parameters. Detections above $5 \sigma$ in the 1.4\,GHz ATLAS image of the field, convolved to the same resolution as the GLEAM image, 
are shown as blue ellipses of the same shapes as their fitted Gaussian parameters. The solid black contour indicates the boundary of the
ATLAS CDFS mosaic.}
\label{fig:GLEAM-CDFS}
\end{figure*}

The resolution of the GLEAM mosaic is 130\,arcsec and that of the ATLAS image is 16.3 by 6.8\,arcsec.
We convolved the ATLAS image to the same resolution as the GLEAM mosaic.
We then ran the source finder \textsc{Aegean} 
\citep{2012MNRAS.422.1812H} on both images using a $5 \sigma$ detection threshold; 133 sources were detected in the ATLAS 
image and 36 in the corresponding region of the GLEAM image.
Figure~\ref{fig:GLEAM-CDFS} shows the GLEAM mosaic of the CDFS, overlayed with the positions of the ATLAS and GLEAM sources. 

All 36 GLEAM sources have a counterpart in ATLAS within 90\,arcsec. While the largest offset between 
ATLAS and GLEAM positions is 90\,arcsec, the second largest offset is only 31\,arcsec; the median 
offset is 9~arcsec. Examination of the original ATLAS image shows that the largest offset is caused by two 
sources being blended in the low-resolution ATLAS and GLEAM images: the ATLAS position lies closer to the 
western source because it is brighter at 1.4\,GHz and the GLEAM position lies closer to the eastern source 
because it is brighter at 150\,MHz.

For many of the ATLAS sources with no counterpart in GLEAM, examination of the GLEAM image shows a weak source
at the ATLAS position detected at the $\sim 3 \sigma$ level. Since all GLEAM sources above $5 \sigma$ are detected in 
ATLAS, we conclude that away from the Galaxy and other extremely bright sources, sources can be 
reliably detected close to a $5\sigma$ detection limit in our GLEAM mosaics.

\section{Sensitivity}
\label{sec:sens}
The MWA is designed for its system temperature to be sky noise dominated over the frequency ranges covered by GLEAM \citep{2013PASA...30....7T}.
As such, the expected thermal noise in a snapshot image depends on the region of sky being observed and, to a lesser extent, the beam pointing used during the observation.

Using image-based linear mosaicing, the thermal noise in the final mosaic is reduced by the weighted contribution of images contributing to the final mosaic.
In the ideal case, the noise is reduced as $\sqrt{N}$ for $N$ images contributing to a given region of the mosaic.
In general, the images contributing to the mosaic are variance weighted by the primary beam \citep{1999ASPC..180..401H}.
For GLEAM, this means that the contribution to the final mosaic from a particular region of sky varies as it drifts through the primary beam.

We calculated the expected thermal noise properties of GLEAM mosaics based on noise-only simulations that match the GLEAM observing strategy.
We used a single fiducial system temperature ($T_{f}=200$\,K) for two image weighting scenarios, as parametrised by the `robust' parameter.
More naturally weighted snapshots (robust=1), which are likely to be used for imaging diffuse emission, have better theoretical sensitivity than more uniformly weighted ones (robust=$-1$), which are more likely to be used for imaging compact sources at the highest resolution.

To calculate the thermal noise in a mosaic we simulate 2 hours of noise-only GLEAM observations for each frequency and declination using \textsc{Miriad}'s `uvgen' task using $T_f$ as the system temperature and matching the duration, duty cycle, bandwidth (including the reduction in usable bandwidth due to commonly flagged channels) and frequency setup of GLEAM.
Each 2\,minute snapshot is imaged with full bandwidth synthesis and weighted according to the primary beam model \citep{2015RaSc...50...52S} for the frequency and declination and accumulated into the mosaic using the `regrid' task.
Likewise, the primary beam `weights' are accumulated into a weight mosaic for each snapshot.
After accumulating all snapshots, the mosaic image is divided by the accumulated weight image.

The noise in the mosaic was measured in approximately $5\times5$ degrees of the inner part of the mosaic using the `imstat' task, and the results for all frequencies and declinations are shown in Table \ref{tab:sensitivity} for both robust parameters.

Clearly the fiducial system temperature $T_f$ is not applicable for most GLEAM frequencies and pointings, hence the values in Table \ref{tab:sensitivity} must be scaled by the appropriate system temperature for region of sky and frequency.

\begin{figure*}
\centering
\hspace{-10mm}
    \begin{subfigure}[b]{0.5\textwidth}
                \label{fig:Tsys_LST_88}
                \includegraphics[width=\textwidth]{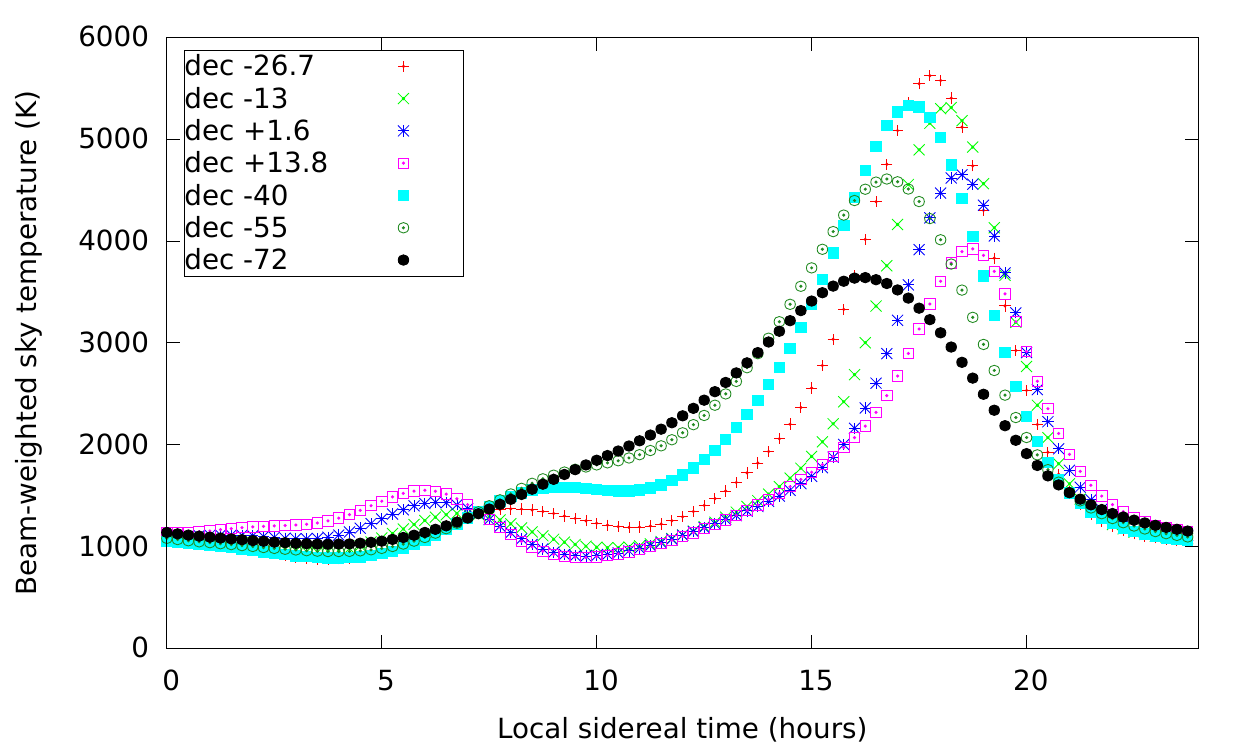}
		\caption{88 MHz}
    \end{subfigure}
    \begin{subfigure}[b]{0.5\textwidth}
                \label{fig:Tsys_LST_118}
                \includegraphics[width=\textwidth]{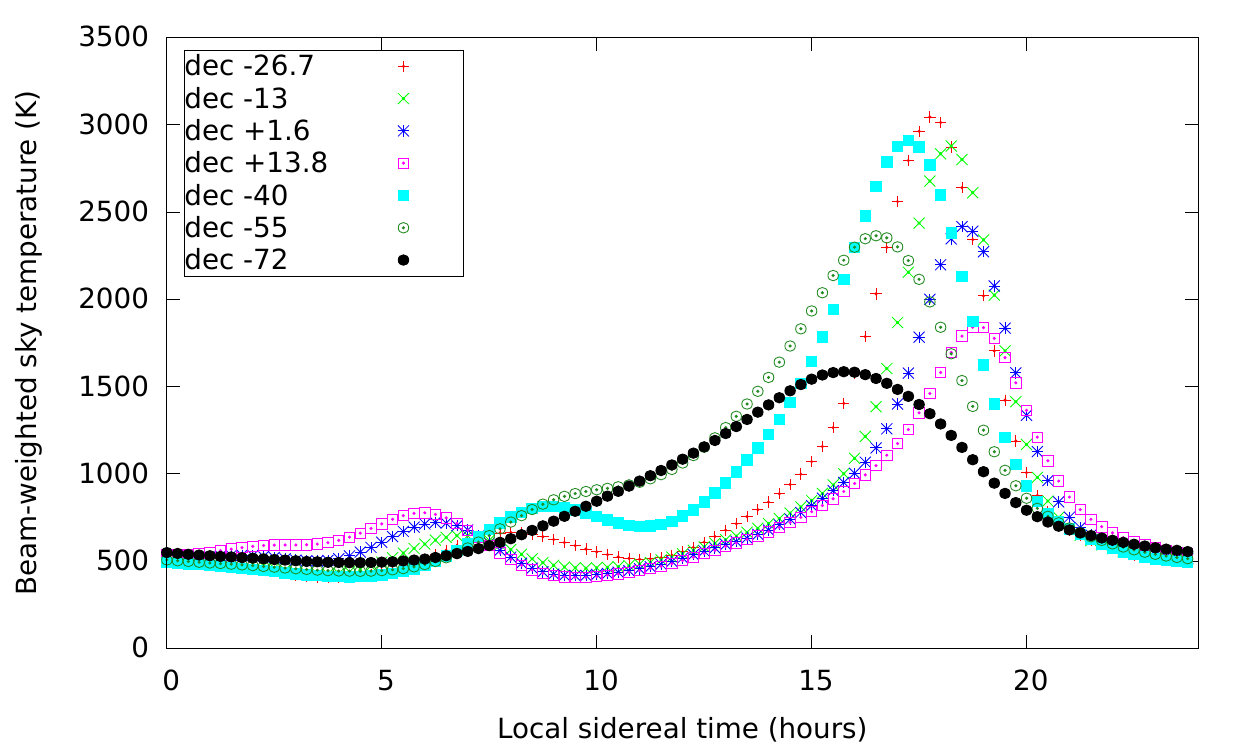}
		\caption{118 MHz}
    \end{subfigure}

\hspace{-10mm}
    \begin{subfigure}[b]{0.5\textwidth}
                \label{fig:Tsys_LST_154}
                \includegraphics[width=\textwidth]{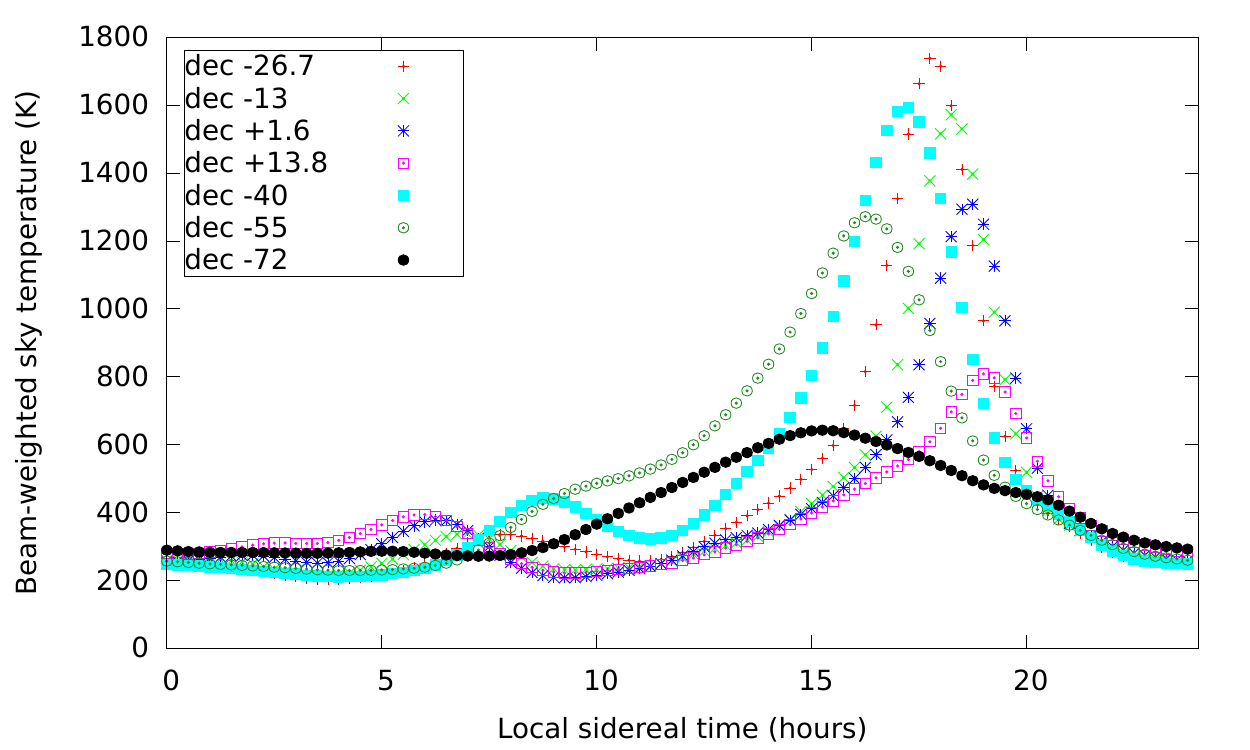}
		\caption{154 MHz}
    \end{subfigure}
    \begin{subfigure}[b]{0.5\textwidth}
                \label{fig:Tsys_LST_185}
                \includegraphics[width=\textwidth]{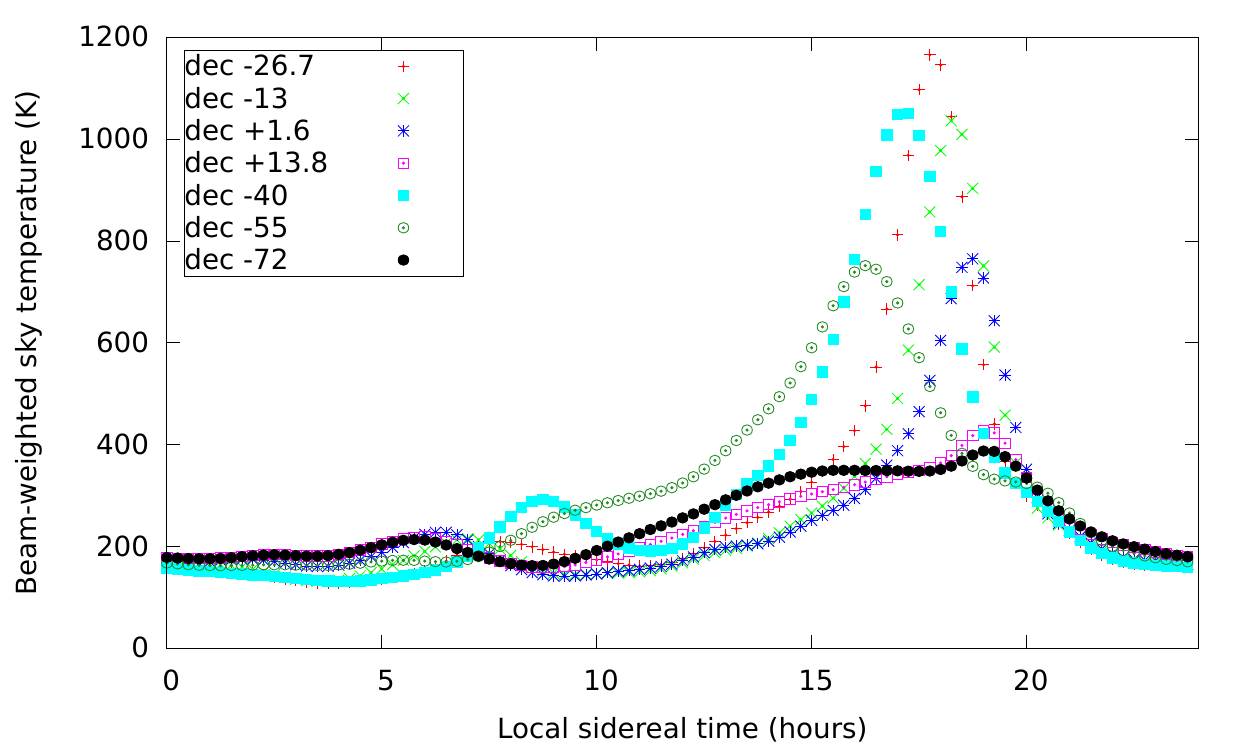}
		\caption{185 MHz}
    \end{subfigure}
    \begin{subfigure}[b]{0.5\textwidth}
                \label{fig:Tsys_LST_216}
                \includegraphics[width=\textwidth]{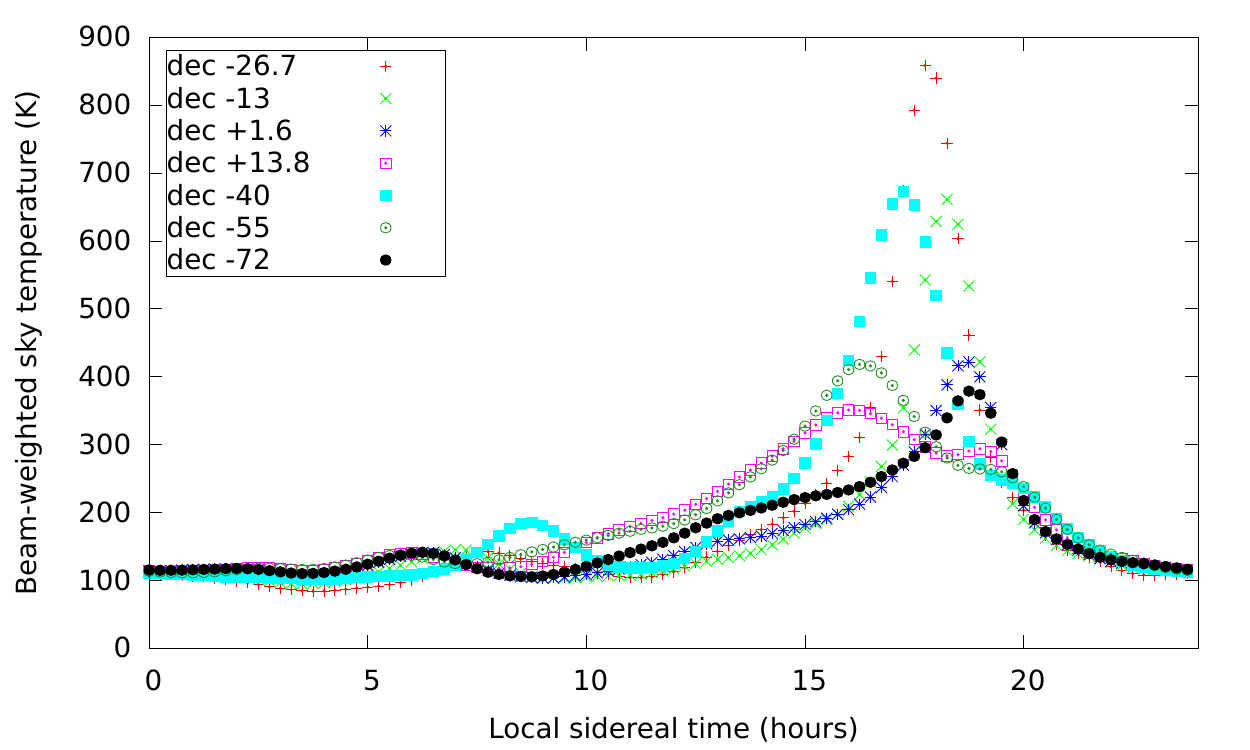}
		\caption{216 MHz}
    \end{subfigure}
    \caption{Beam-weighted sky average temperature for the seven declinations used by GLEAM depending on LST. Panels show the five central frequencies used by GLEAM: 88, 118, 154, 185 and 216\,MHz.}
    \label{fig:Tsys_vs_LST}
\end{figure*}

To enable this scaling, we calculated the expected beam-weighted average sky temperature over the range of frequencies, pointings and LSTs relevant to GLEAM (Figure \ref{fig:Tsys_vs_LST}).
The correct theoretical thermal noise in an image can then be derived by calculating the correct $T_{sys}$ for that image by scaling by the ratio of the average sky temperature ($T_{true}$) for that frequency/LST/pointing, plus receiver noise ($T_R \approx 50$\,K),  to the fiducial value, i.e. by $(T_{true}+T_R)/T_{f}$.
For example, at 154\,MHz the `cold' extragalactic sky will generate a system temperature of approximately 300\,K (Figure \ref{fig:Tsys_vs_LST}) hence the theoretical thermal noise for a uniformly weighted mosaic of that region of the sky will be $(300/200) \times 2.1$\,mJy $=3.2$\,mJy\,beam$^{-1}$ (near the zenith).

\begin{table*}
  \caption{GLEAM expected thermal noise sensitivity (mJy\,beam$^{-1}$) from a 2 hour mosaic, assuming fiducial system temperature $T_f = 200$\,K. Columns are the frequency in MHz, rows are the declination in degrees.}
  \centering
  \begin{subtable}{0.4\textwidth}
    \begin{tabular}{ l | c c c c c }
    \hline \hline
      & 88 &  118 &  154 & 185 &  216 \\ \hline
$-72$  & 3.2 & 3.0 & 2.3 & 3.8 & 5.1 \\
$-55$  & 2.1 & 2.1 & 1.9 & 2.3 & 4.3 \\
$-40.5$  & 1.8 & 1.9 & 2.0 & 2.1 & 3.0 \\
$-26.7$  & 1.7 & 1.9 & 2.1 & 2.2 & 2.4 \\
$-13$  & 1.8 & 1.9 & 2.1 & 2.3 & 3.4 \\
$+1.6$  & 2.1 & 2.2 & 2.2 & 2.8 & 5.3 \\
$+18.3$  & 3.4 & 3.4 & 2.9 & 5.2 & 7.1 \\
\hline \hline
    \end{tabular}
    \caption{Robust -1}
  \end{subtable}
  \begin{subtable}{0.4\textwidth}
    \begin{tabular}{ l | c c c c c }
    \hline \hline
      & 88 &  118 &  154 & 185 &  216 \\ \hline
$-72$  & 1.5 & 1.4 & 1.1 & 1.8 & 2.4 \\
$-55$  & 1.0 & 1.0 & 0.9 & 1.1 & 2.1 \\
$-40.5$  & 0.9 & 0.9 & 1.0 & 1.0 & 1.5 \\
$-26.7$  & 0.9 & 0.9 & 1.0 & 1.1 & 1.1 \\
$-13$  & 0.9 & 1.0 & 1.1 & 1.1 & 1.6 \\
$+1.6$  & 1.1 & 1.2 & 1.1 & 1.4 & 2.7 \\
$+18.3$  & 1.7 & 1.7 & 1.4 & 2.5 & 3.3 \\
\hline \hline
    \end{tabular}
    \caption{Robust +1}
  \end{subtable}
\label{tab:sensitivity}
\end{table*}

\subsection{Confusion and excess image noise in Stokes I}
\label{sec:confusion}

The MWA's synthesised beam is large enough that classical confusion should become a limiting factor in wide bandwidth mosaics or long duration synthesis images, compared to thermal noise.
It is difficult to precisely estimate the classical confusion because the differential source counts at low frequencies between 1 and 100\,mJy are currently not well known.
To estimate the classical confusion we model the differential source counts in this flux density range as $n(S) = k S^{\gamma}$.
Based on \citet{1991PhDT.......241W}, who measured the 327\,MHz counts down to $\sim1$\,mJy using the WSRT, we estimate $k=4000$ and $\gamma=-1.6$ at 327\,MHz.
Assuming the spectral index of sources is $-0.7$, $k$ can be scaled to frequency $\nu$\,MHz by the factor $(\nu/327)^{-0.7(1+\gamma)}$.
Using this differential source count model scaled to 154\,MHz and following \citet{1974ApJ...188..279C} using a signal-to-noise threshold of 6, the $1\sigma$ classical confusion for an image with synthesised beam size 2.4\,arcmin (i.e. a uniformly weighted image made at 154\,MHz) is approximately 2\,mJy.
There is substantial uncertainty in this estimate because the overall scaling is uncertain at the $\sim25$\% level and because the slope of the source count function is known to change at mJy flux densities at higher frequencies \citep[e.g.][]{1990ASPC...10..389W}.
For example, a small change in $\gamma$ from -1.6 to -1.8 around mJy flux densities will triple the estimated classical confusion.
GLEAM and other contemporary low-frequency surveys will make a substantial contribution to better understand the low-frequency source counts.

Full synthesis 2\,minute Stokes I snapshot images have a theoretical thermal noise of 5\,mJy for $T_{sys}=200$\,K using robust~$-1$.
Despite this, such images typically contain $\sim20-30$\,mJy background noise and the background between instrumental polarisation (XX and YY) images tends to be correlated.
The reason for the excess background is still under investigation, but is most likely due to a combination of classical and sidelobe confusion (sidelobe confusion is the extra background variance due to the combined sidelobes of all the faint unsubtracted sources within the primary beam).
Sidelobe confusion is more pronounced for the MWA than for radio telescopes with larger antennas due to the MWA's huge field-of-view.
A detailed analysis of the impact of sidelobe confusion to MWA snapshot and long duration synthesis images is underway (Wayth et al., in prep).
For the time being the excess background, which is only found in Stokes I, is treated simply as excess noise which integrates down with time and bandwidth.

\section{GLEAM outputs}
\label{sec:outputs}

We plan to process GLEAM data and release data products in stages. The expected data products from GLEAM include: an extragalactic Stokes I compact source catalogue; a full polarisation compact source catalogue; maps of diffuse extragalactic polarised foreground; and maps of the diffuse emission (both Stokes I and polarised) in the Galactic plane.
Most sources will have at least 5 independent continuum flux density measurements made over GLEAM's frequency range and further subdivision of the frequency range is possible.

\subsection{Extragalactic sources}
\label{sec:output_extragal}
GLEAM covers 7.5\,sr of extragalactic ($|b|>10$) sky.
Assuming a $6\sigma$ source detection threshold of 120\,mJy in the 180\,MHz frequency band, we again model the differential source counts, $n(S)$, between 0.1 and 1.0\,Jy as a power-law based on the general properties of source counts at similar frequencies \citep[e.g.][]{1991PhDT.......241W,2013A&A...549A..55W}.
Using\footnote{The normalisation and slope of the model for $n(S)$ here differs from \S\ref{sec:confusion} because we are using a different part of the source count curve.}
$n(S) = 3600 S^{-1.78}$ and a detection threshold of 120\,mJy, we estimate GLEAM will detect approximately 19,500 sources\,$\mathrm{sr}^{-1}$ or 150,000 sources in the extragalactic sky visible to the MWA.
Bright residual structure extending far from the Galactic plane and regions around bright compact sources may have higher than typical background noise, which will affect source detection locally.
By comparison, the MWA Commissioning Survey \citep[`MWACS',][]{2014MWACS} found 7,540 sources\,$\mathrm{sr}^{-1}$ with a detection threshold of 200\,mJy or greater, depending on the local noise properties of the mosaic.

\subsection{Comparison to other southern hemisphere surveys}
There have been relatively few sky surveys in the southern hemisphere compared to the northern, especially at low frequencies.
Surveys below 1\,GHz that cover a substantial fraction of the southern hemisphere are listed in Table \ref{tab:surveys_summary}.
We also note the Culgoora array observed selected sources at 80 and 160\,MHz \citep{1995AuJPh..48..143S}, but did not perform a blind survey.

Table \ref{tab:surveys_summary} shows that GLEAM is comparable in angular resolution and sky coverage to the all extragalactic sky Molonglo Reference Catalogue (MRC), but will be an order of magnitude more sensitive with full polarisation and sensitivity to very large structures.
GLEAM will complement the TGSS and VLSS surveys with similar sensitivity, but slightly poorer resolution, filling in the entire sky south of $\delta=+25$.
GLEAM will also form an excellent complementary dataset to the LOFAR MSSS \citep{Heald15}, which will have similar sensitivity and angular resolution in the northern hemisphere.
GLEAM will be unparalleled in its ability to image large, low surface brightness structures, both Galactic and extragalactic, in full polarisation.
The strengths of MWA's surface brightness sensitivity have already been made evident by the serendipitous discovery of a relic radio galaxy in early MWA data \citep{2015MNRAS.447.2468H}.

\begin{table*}
  \caption{Summary of radio surveys below 1\,GHz substantially covering the southern hemisphere}
  \centering
  \begin{threeparttable}
  \begin{tabular}{ l | c | c | c | c | c }
  \hline \hline
    		        & Freq  & Resolution	& Max size  & 		            & Stokes I	     \\
 Survey 	        & (MHz) & (arcmin) 	    & (arcmin)  & Coverage 	        & cutoff (Jy)	  \\ \hline
 MRC\tnote{a}  	    &  408  & $2.6\times2.9\,\mathrm{sec}(\delta +35.5^{\circ})$	& $\sim30$ & $+18.5 > \delta > -85$, $|b|>3$	& 0.7 \\
 SUMSS\tnote{b} 	& 843   & $0.75\times0.75\,\mathrm{cosec}|\delta|$   & $163$ & $\delta < -30$	& 0.006 - 0.01   \\
 VLSS(r)\tnote{c}  & 74	    & 1.25 	        & $\sim23$\tnote{*} & $\delta > -40$	& $\sim 0.5$	    \\
 TGSS\tnote{d}	    & 150   & 0.33 	        &           & $\delta > -30$	& $\sim 0.03$	  \\
 PAPER32\tnote{e}  & 145   & 26            & $\sim300$ & $\delta < 10$    & 10               \\
 MSH\tnote{f}      & 86    & 50            &  n/a    & $\delta < 10$      &  20 \\
 GLEAM	& 72-231& $2.5 \times 2.2\sec(\delta+26.7^{\circ})$\tnote{\textdagger} & $\sim 600$ & $\delta < +25$	&  $\sim 0.1$\tnote{\textdagger} \\	\hline \hline
\end{tabular}
\begin{tablenotes}
    \item[*] Assuming $150\lambda$ is the shortest effective baseline for the VLA B array at 74\,MHz.
    \item[\textdagger] At 154\,MHz.
    \item[a] \citet{1981MNRAS.194..693L}.
    \item[b] \citet{1999AJ....117.1578B,2003MNRAS.342.1117M}. See also MGPS-2 \citep{2007MNRAS.382..382M} for sources with $|b| < 10$.
    \item[c] \citet{2007AJ....134.1245C,2014MNRAS.440..327L}.
    \item[d] \url{http://tgss.ncra.tifr.res.in }.
    \item[e] \citet{2011ApJ...734L..34J}.
    \item[f] \citet{1958AuJPh..11..360M,1960AuJPh..13..676M,1961AuJPh..14..497M}.
\end{tablenotes}
\end{threeparttable}
\label{tab:surveys_summary}
\end{table*}

\section{Conclusion}
\label{sec:conclusion}

We have presented the observing strategy, data processing strategy and theoretical sensitivity of the MWA GLEAM survey.
GLEAM covers the entire radio sky south of declination $+25^{\circ}$ between 72 and 231\,MHz.
GLEAM aims to leave a significant legacy dataset for the MWA and GLEAM data are being used for many Galactic, extragalactic and time domain science programs.
Data products will include a compact source catalogue and maps of the diffuse Galactic and extragalactic sky, both in Stokes I and in polarisation.
The Stokes I compact source catalogue, in particular, will have an order of magnitude improvement in sensitivity compared to the MRC.
Looking towards the SKA era, GLEAM will provide a foundation sky model that can be used in preparation for SKA-low key science programs.

GLEAM data are unparalleled in areas where the MWA's strengths lie, including very wide field-of-view, full polarisation, high surface brightness sensitivity at large angular scales and broad frequency coverage.
The broad frequency coverage and fine frequency resolution of GLEAM provide a large lever arm for both polarisation studies using RM synthesis and to measure spectral indices.

We showed example snapshot images from GLEAM and discussed practical issues associated with forming mosaics from GLEAM including calibration, ionospheric and primary beam polarisation effects, as well as the effect of strong sources in the primary beam sidelobes.
We discussed how confusion impacts the background noise level in Stokes I images, which is a consequence the MWA's very large field-of-view.

Finally, we calculated the theoretical thermal noise sensitivity for GLEAM mosaics and showed how the expected thermal noise depends on the region of the sky being imaged, the observing frequency and the image weighting scheme.


\begin{acknowledgements}
This scientific work makes use of the Murchison Radio-astronomy Observatory, operated by CSIRO. We acknowledge the Wajarri Yamatji people as the traditional owners of the Observatory site. Support for the MWA comes from the U.S. National Science Foundation (grants AST-0457585, PHY-0835713, CAREER-0847753, and AST-0908884), the Australian Research Council (LIEF grants LE0775621 and LE0882938), the U.S. Air Force Office of Scientific Research (grant FA9550-0510247), and the Centre for All-sky Astrophysics (an Australian Research Council Centre of Excellence funded by grant CE110001020). Support is also provided by the Smithsonian Astrophysical Observatory, the MIT School of Science, the Raman Research Institute, the Australian National University, and the Victoria University of Wellington (via grant MED-E1799 from the New Zealand Ministry of Economic Development and an IBM Shared University Research Grant). The Australian Federal government provides additional support via the Commonwealth Scientific and Industrial Research Organisation (CSIRO), National Collaborative Research Infrastructure Strategy, Education Investment Fund, and the Australia India Strategic Research Fund, and Astronomy Australia Limited, under contract to Curtin University. This work was supported by resources provided by the Pawsey Supercomputing Centre with funding from the Australian Government and the Government of Western Australia. We acknowledge the iVEC Petabyte Data Store, the Initiative in Innovative Computing and the CUDA Center for Excellence sponsored by NVIDIA at Harvard University, and the International Centre for Radio Astronomy Research (ICRAR), a Joint Venture of Curtin University and The University of Western Australia, funded by the Western Australian State government.
\end{acknowledgements}

\newcommand{\pasa}{PASA}
\newcommand{\apj}{ApJ}
\newcommand{\aj}{AJ}
\newcommand{\apjl}{ApJ}
\newcommand{\aap}{A\&A}
\newcommand{\mnras}{MNRAS}
\newcommand{\aaps}{A\&AS}
\newcommand{\pasp}{PASP}
\newcommand{\physrep}{PhR}
\newcommand{\araa}{ARA\&A}
\bibliographystyle{apj}
\bibliography{refs}

\end{document}